\documentstyle[12pt,fullpage]{article}

\begin{document}
\parskip=10pt
\begin{flushright}
TAUP-2076-93
\end{flushright}
\vskip 3 true cm

\begin{center}
\Large
{\bf On Feynman's Approach to the Foundations of Gauge Theory}
\normalsize

M. C. Land$^{\ddag}$, N. Shnerb$^{\S}$, and L. P. Horwitz$^{\ddag\S}$

$^{\ddag}$School of Physics and Astronomy  \\
Raymond and Beverly Sackler Faculty of Exact Sciences  \\
Tel Aviv University, Ramat Aviv, Israel

$^{\S}$Department of Physics \\ Bar-Ilan University \\
Ramat Gan, Israel
\end{center}
\vskip .5 true cm
\baselineskip 7mm
\parindent=0cm

\begin{abstract}
\parindent=0cm
\parskip=10pt
In 1948, Feynman showed Dyson how the Lorentz force law and
homogeneous Maxwell equations could be derived from commutation
relations among Euclidean coordinates and velocities, without
reference to an action or variational principle.  When Dyson
published the work in 1990, several authors noted that the
derived equations have only Galilean symmetry and so are not
actually the Maxwell theory.  In particular, Hojman and Shepley
proved that the existence of commutation relations is a strong
assumption, sufficient to determine the corresponding action,
which for Feynman's derivation is of Newtonian form.  In a
recent paper, Tanimura generalized Feynman's derivation to a
Lorentz covariant form with scalar evolution parameter, and
obtained an expression for the Lorentz force which appears to be
consistent with relativistic kinematics and relates the force to
the Maxwell field in the usual manner.  However, Tanimura's
derivation does not lead to the usual Maxwell theory either,
because the force equation depends on a fifth ({\it scalar})
electromagnetic potential, and the invariant evolution parameter
cannot be consistently identified with the proper time of the
particle motion.  Moreover, the derivation cannot be made
reparameterization invariant; the scalar potential
causes violations of the mass-shell constraint which this
invariance should guarantee.

In this paper, we examine Tanimura's derivation in the framework
of the proper time method in relativistic mechanics, and use the
technique of Hojman and Shepley to study the unconstrained
commutation relations.  We show that Tanimura's result then
corresponds to the five-dimensional electromagnetic theory
previously derived from a Stueckelberg-type quantum theory in
which one gauges the invariant parameter in the proper time
method.  This theory provides the final step in Feynman's
program of deriving the Maxwell theory from commutation
relations; the Maxwell theory emerges as the ``correlation limit''
of a more general gauge theory, in which it is properly
contained.

%
\end{abstract}
\section{Introduction}
\setcounter{equation}{0}

In 1990, Dyson \cite{Dyson} published a derivation, originally
due to Feynman, of the Lorentz force law in the form
\begin{equation}
m \frac{d^2 x^i }{dt^2} = E^i(t,x) + \epsilon^{ijk} \frac{dx_j}{dt}
H_k (t,x),
\label{eqn:1.1}
\end{equation}
and of the homogeneous Maxwell equations
\begin{equation}
\nabla \cdot {\bf H} = 0 \qquad \nabla \times {\bf E} +
\frac{\partial}{\partial t} {\bf H} =0,
\label{eqn:1.2}
\end{equation}
which assumes only the canonical commutation relations among
Euclidean position and velocity,
\begin{eqnarray}
[ \: x^i, x^j \: ] &=& 0,
\label{eqn:1.3}
\\
m \, [ \: x^i, \dot x^j \: ] &=& i \hbar \, \delta^{ij},
\label{eqn:1.4}
\end{eqnarray}
and Newton's second law
\begin{equation}
m \: \ddot x^i = F^i (t,x,\dot x),
\label{eqn:1.5}
\end{equation}
where $\dot x^i = dx^i / dt$ and $i,j=1,2,3$.  Feynman's program
of deriving the Maxwell theory without reference to either
canonical momenta or a Lagrangian was based on the hypothesis
that commutation relations between $x$ and $\dot x$ form a
fundamental basis for mechanics, but constitute a weaker set of
initial assumptions, and might therefore lead to a more general
theory.  Dyson explained that Feynman had shown him this
derivation in October 1948 but never published it because, ``He
was exploring possible alternatives to the standard theory,''
which this proof, by producing the usual equations, failed to
provide.

Dyson's publication of Feynman's derivation provoked a debate in
the literature \cite{comments,vaidya,H-S,Hughes}, with several
papers challenging
the identification of the derived equations as Maxwell's theory.
These authors argue that although the Lorentz covariance of
(\ref{eqn:1.2}) may be assumed {\it a posteriori}, Lorentz
covariance of the inhomogeneous Maxwell equations conflicts with
the Euclidean assumptions of the ``proof.''  In fact, as pointed
out by \cite{vaidya} the homogeneous equations may be regarded as
Galilean or Lorentz covariant; however, the inhomogeneous Maxwell
equations are not Galilean covariant and the Lorentz force
equations are not Lorentz covariant.

This conflict of symmetries was demonstrated in a stronger
fashion by Hojman and Shepley \cite{H-S} and by Hughes
\cite{Hughes} who place Feynman's program in the context of the
inverse problem of the calculus of variations and demonstrate
that conditions (\ref{eqn:1.3}) and (\ref{eqn:1.4}) are
sufficient to establish the self-adjointness of the differential
equations (\ref{eqn:1.5}).  Given self-adjointness, it is well
known \cite{Santilli} that the right hand side of (\ref{eqn:1.1})
is the most general admissible form for $F^i(t,x,\dot x)$, and
that this system of differential equations is equivalent to the
Lagrangian formulation
\begin{equation}
L = \frac{1}{2}m \delta_{ij} \dot x^i \dot x^j +
\delta_{ij} \dot x^i A^j (t,x) + A_0 (t,x)
\label{eqn:1.6}
\end{equation}
where
\begin{equation}
H =\nabla \times {\bf A} \qquad
{\bf E} = -\left(\frac{\partial}{\partial t} {\bf A} +
\nabla A_0 \right).
\label{eqn:1.7}
\end{equation}
The potentials $A_0$ and ${\bf A}$ must exist by virtue of
(\ref{eqn:1.2}).  Hence, a Lagrangian and well-defined canonical
momenta exist, and these are of Galilean form.  Furthermore, the
inhomogeneous equations of Maxwell form, which can be obtained by
adding kinetic terms for the fields, would not be Lorentz
covariant since the source current is not Lorentz covariant.
Therefore, Feynman's argument essentially re-derives the
conditions on velocity dependent forces in nonrelativistic
mechanics.  These conditions are essential to the self-adjointness
of the differential equations (a consequence of the commutation
relations), but do not reveal the full dynamical structure
implied by the commutation relations.

In an attempt to achieve a properly relativistic version of
Feynman's derivation, Tanimura \cite{Tanimura} has presented a
modified argument which maintains manifest Lorentz covariance
(in $d$-dimensional spacetime) throughout.  Employing the
approach which is the basis of the ``proper time formalism''
\cite{Fock,Nambu,Stueckelberg,Feynman,Schwinger}, Tanimura
assumes the commutation relations
\begin{eqnarray}
[ \: x^\mu, x^\nu \: ] &=& 0,
\label{eqn:1.8}
\\
m \, [ \: x^\mu, \dot x^\nu \: ] &=& - i \hbar \, g^{\mu \nu} (x),
\label{eqn:1.9}
\end{eqnarray}
and the force law
\begin{equation}
m \: \ddot x^\mu = F^\mu (x,\dot x),
\label{eqn:1.10}
\end{equation}
where $\mu,\nu = 0,\cdots,d-1$, and $\dot x = \frac{dx}{d\tau} $,
where $\tau$ is a Lorentz invariant parameter.  In analogy to
Feynman's derivation, he obtains expressions which are formally
similar to the usual covariant form of the Lorentz force in
curved space
\begin{equation}
m \ddot x^\mu = F^\mu(x,\dot x) = G^\mu(x) +
F^\mu_{\;\;\;\nu}(x) \, \dot x^\nu -
m \Gamma^\mu_{\;\:\nu \rho} \,
\dot x^\nu \, \dot x^\rho ,
\label{eqn:1.11}
\end{equation}
where $\Gamma^\mu_{\;\:\nu \rho}$ is the Levi-Civita connection
\begin{equation}
\Gamma_{\mu \nu \rho} = \frac12 (
\partial_\rho g_{\mu \nu}+\partial_\nu  g_{\mu \rho}-
\partial_\mu  g_{\nu \rho})
\label{eqn:1.12}
\end{equation}
(so that the terms proportional to $m$ are the covariant
derivative of the velocity),
and the homogeneous Maxwell equations,
\begin{equation}
\partial_\mu F_{\nu \rho} + \partial_\nu F_{\rho \mu} +
\partial_\rho F_{\mu \nu} =0.
\label{eqn:1.13}
\end{equation}
The vector field $G^\mu (x)$ satisfies
\begin{equation}
\partial_\mu G_\nu - \partial_\nu G_\mu = 0.
\label{eqn:1.14}
\end{equation}
{}From (\ref{eqn:1.13}) and (\ref{eqn:1.14}) one sees that
$F_{\mu\nu}$ and $G_\mu$ may be derived from a $d$-vector and
a scalar potential, respectively.

However, the expressions which Tanimura derives are not the
Maxwell theory, either.  In addition to the usual antisymmetric
second rank tensor $F^{\mu\nu}$, the electromagnetic field
includes the ``new'' $d$-vector field $G_\mu$, and both of these
fields contribute to the Lorentz force.  But, $F^{\mu\nu}$ and
$G_\mu$ are completely decoupled in the field equations, so that
while $G^\mu(x)$ could act nontrivially on particle motions, no
interaction with the usual electromagnetic field is available to
control the scalar potential.  Moreover, Tanimura finds that the
variable $\tau$ introduced to parameterize the world lines cannot
be taken to be the proper time of the motion (the proper time
constraint $\dot x^\mu \dot x_\mu = 1$ and (\ref{eqn:1.9}) imply
$\dot x^\mu =0$) and must be treated as a ``new'' Lorentz scalar,
independent of the phase space coordinates.  Similarly, the
derivation is not invariant under general reparameterization ---
only affine transformations of $\tau$ preserve the structure of
the equations.  Since these differences from standard Maxwell
electrodynamics originate in the framework of the proper time method,
they merit further study and explanation.

In this paper, we shall show that the appearance of a ``new''
scalar potential and a ``new'' scalar time are necessary and
related consequences of the unconstrained commutation relations
assumed in the formulation of the problem.  By assuming, in
(\ref{eqn:1.8}) and (\ref{eqn:1.9}), that the $2d$ components of
position and velocity are independent, Tanimura defines an
unconstrained canonical problem in $d+1$-dimensions with an
implicit symplectic structure analogous to that of the Newtonian
problem posed by Feynman.  Applying the technique of Hojman and
Shepley to the relativistic case, we shall construct the
equivalent Lagrangian and Hamiltonian formulations of this
canonical problem, which turn out to be naturally Lorentz
and gauge invariant.  Since the form of the
Lagrangian is determined by the commutation relations, any
constraints imposed on the system subsequently must be consistent
with (\ref{eqn:1.8}) and (\ref{eqn:1.9}).  Thus, if we require
that the parameter $\tau$ be {\it a priori} proportional to the
proper time, then the relation
\begin{equation}
\dot x^\mu \dot x_\mu = (ds/d\tau)^2 = {\rm constant}
\label{eqn:1.15}
\end{equation}
would be constitute a constraint on the velocities, contradicting
the assumptions of the problem.  Similarly, the validity of
operations of the form
\begin{equation}
\frac{d}{d\tau} [x^\mu (\tau) , \dot x^\mu (\tau) ] =
[\dot x^\mu (\tau) , \dot x^\mu (\tau) ] +
[x^\mu (\tau) , \ddot x^\mu (\tau) ]
\label{eqn:1.16}
\end{equation}
(a central step in the Feynman and Tanimura derivations) becomes
questionable when $\tau$ is not completely independent of the
coordinates and velocities.  For the same reasons, general
reparameterization invariance is inconsistent with equations
(\ref{eqn:1.8}) and (\ref{eqn:1.9}), because such invariance
corresponds to a constraint on the phase space \cite{hartle}.
Following Hojman and Shepley, we shall be led to an action in
which reparameterization invariance is clearly absent. We will
further demonstrate that the ``new'' gauge degree of freedom,
which appears through the field $G^\mu$, acts as a compensation
field for the dependence of local gauge transformations on the
parameter $\tau$, the ``new'' dimension independent of the
spacetime variables.  So, just as in the Newtonian problem, it is
the dimension of the dynamical problem which determines the
number of gauge degrees of freedom.  Loosely speaking, one may
think of the proper time constraint in the conventional Maxwell
theory as removing the additional gauge degree of freedom (we
will sharpen this point below).

The problem of implementing the canonical commutation relations
\begin{equation}
[ x^\mu , p_\nu ] = - i \hbar \, \delta^\mu \ _\nu
\label{eqn:1.17}
\end{equation}
on the 8 dimensional phase space coordinates is an old one, whose
principal difficulties are clearly seen in the course of
Tanimura's derivation.  Since the observed time is a coordinate in
Minkowski space, it may not be used as a parameter of system
evolution (in light of the Feynman-Stueckelberg
\cite{interpretation} interpretation of negative energy states as
propagating backward in time, the motion need not even be monotonic
in $t$).  But
if one parameterizes the coordinates with a Lorentz invariant
time (which by the arguments of the previous paragraph, may not
be identified with the proper time of the phase space) and
maintains the notion of a definite particle mass, then one must
adequately handle the proper time (or mass-shell) constraint.  In
the original formulation of the so-called proper time method
\cite{Nambu,Stueckelberg,Feynman,Schwinger,lusanna,llosa}, the
constraint was applied {\it a posteriori} to the solutions of the
unconstrained problem, which is permissible as long as the
interactions preserve the mass-shell dynamically (we shall examine
this point below).  In more recent formulations
\cite{hartle,t-h,Mannheim,Sathiapalan},
the constrained theory is rewritten in a form in which a Lagrange
multiplier enforces the constraint dynamically.  In both of these
methods, one associates with the system a manifestly covariant
canonical mechanics with invariant evolution parameter,
permitting the application of techniques from nonrelativistic
mechanics.  In addition, as pointed out by Schwinger
\cite{Schwinger}, since the physical interactions are independent
of the evolution parameter, the proper time method preserves the
symmetries of the system.

In a different approach to the proper time formalism, introduced
by Horwitz and Piron \cite{H-P}, the invariant evolution
parameter $\tau$ is regarded as a true physical time, playing the
role of the Newtonian time in nonrelativistic mechanics.  One is
then led to a symplectic mechanics in which manifest Poincar\'e
covariance plays the role of Galilean covariance in Newtonian
mechanics (see also \cite{others}).
In this theory, the proper time relation
(\ref{eqn:1.15}) is not a constraint at all.  The value of $\dot
x^\mu \dot x_\mu$ is a
dynamical quantity; it may be constant only for appropriate (for
example, electromagnetic \cite{Stueckelberg}) interactions.  The
relaxation of the constraint permits the consideration of
interactions of a more general type, and in particular allows the
construction of consistent relativistic potential models.  The
Hamiltonian form of this mechanics leads naturally to a
Schr\"odinger type equation, which has been solved for the
relativistic bound state \cite{bound} and scattering
problems \cite{A-S}.
Arguing that local gauge invariance of the Schr\"odinger equation
should include gauge transformations dependent on the evolution
parameter, Sa'ad, Horwitz, and Arshansky \cite{Saad} introduced a
compensation field associated with the invariant time $\tau$.
This new potential leads to an off-shell electromagnetic theory
(so called because the interactions explicitly take the
electromagnetic fields off mass shell), which
nevertheless has the Maxwell theory as its ($\tau$-static)
equilibrium limit \cite{shnerb}.  The quantum field theory of
off-shell electromagnetism has been developed and provides a
basis for the empirical evaluation of this structure
\cite{shnerb,frastai,qft}.

We examine here Tanimura's work in the context of the proper time
method.  We will show that the force and field equations obtained
by Tanimura are just those of the off-shell theory.  By adapting
the techniques of Hojman and Shepley, we will demonstrate that
Tanimura's assumptions lead to the Lagrangian of off-shell
electromagnetism.  Thus, while Tanimura's derivation does not lead
directly to the Maxwell theory, it does lead to a proper
generalization of Maxwell's electromagnetism, which goes over to
the Maxwell theory in a systematic manner, and therefore fulfills
Feynman's program of providing a path from the canonical
commutation relations to Maxwell's theory, as well as a more
general theory.

That Tanimura's derivation leads to a set of equations associated
with an off-shell dynamical theory leads to an interesting
connection between commutation relations and gauge degrees of
freedom.  This connection is of particular importance as the
proper time method is increasingly used with interactions
depending explicitly on the proper time parameter \cite{Kaplan}.
As we shall demonstrate below, such
interactions require that the ``new'' gauge field associated with
the proper time dimension, possess a non-trivial relationship to
the usual gauge fields and are inconsistent with the requirements of
on-shell dynamics.

In Section 2, we review Tanimura's covariant derivation in
curved spacetime, and obtain equations (\ref{eqn:1.11}),
(\ref{eqn:1.13}) and (\ref{eqn:1.14}) from equations
(\ref{eqn:1.8}) and (\ref{eqn:1.9}).  In Section 3, we provide a
brief review of the inverse problem in the calculus of
variations and the work of Hojman and Shepley.  We use these
results to obtain Tanimura's equations from self-adjointness.
In Section 4, we present the theory of off-shell
electromagnetism as the local gauge theory of covariant quantum
mechanics with invariant evolution parameter, and shall show
that it is precisely the theory derived in Sections 2 and 3.  We
then show how the theory differs from Maxwell electrodynamics as
a dynamic theory, but reduces to it in the $\tau$-static limit.
In the context of the off-shell theory, we will discuss the
related issues of constraints and gauge freedom.  In Section 5,
the presentation of Section 3 is generalized to the case of
non-Abelian gauge fields, and the results are compared with
the equations obtained by Tanimura.

\section{The Lorentz Force in Curved Spacetime}
\setcounter{equation}{0}

We now review the
derivation given by Tanimura \cite{Tanimura}
in a pseudo-Riemannian manifold.
In this theory, we treat the event as a point in the
$d$-dimensional curved spacetime whose coordinates
$x^\mu(\tau)$, $(\mu=0,1,\cdots,d-1)$ evolve according to the
invariant parameter $\tau$.  This approach has the advantage that
the conjugate velocities are derivatives of the event coordinates
with respect to $\tau$, a well defined procedure even in curved
spacetime \cite{GR}.

We consider the metric $g_{\mu\nu}(x)$ (which becomes
$\eta_{\mu \nu} =\mbox{diag} (+1,-1,\cdots,-1) $ in flat
space), so that the coordinates and velocities $\dot x^\mu(\tau)$
are assumed to satisfy the commutation relations
\begin{eqnarray}
[x^\mu,x^\nu] &=& 0
\label{eqn:2.1} \\
m[x^\mu,\dot x^\nu] &=& -i\hbar g^{\mu\nu}(x)
\label{eqn:2.2}
\end{eqnarray}
and the equations of motion are
\begin{equation}
m \ddot x^\mu = F^\mu (\tau,x,\dot x).
\label{eqn:2.3}
\end{equation}
We regard $x^\mu (\tau)$, its $\tau$-derivatives and functions of
these quantities as quantum operators in a Heisenberg picture.
The field equations and the Lorentz force may then be
interpreted,
in the Ehrenfest sense, as relations among the expectation values
which correspond to relations among classical quantities.
Generalizing Tanimura, we allow the force $F^\mu (\tau,x,\dot x)$
to be a function of $\tau$.
Equation (\ref{eqn:2.2}) implies that for any function $ q(x)$
\begin{equation}
[\dot x^\mu, q(x)] = {i\hbar \over m}{ \partial q  \over
\partial x_\mu} .
\label{eqn:2.4}
\end{equation}
We differentiate (\ref{eqn:2.2}) with respect to $\tau$
\begin{equation}
m [\dot x^\mu, \dot x^\nu] + m [x^\mu , \ddot x^\nu ] =
-i\hbar \partial_\rho g^{\mu\nu} (x)\dot x^\rho
\label{eqn:2.5}
\end{equation}
and define $W^{\mu\nu} = - W^{\nu\mu}$ by
\begin{equation}
W^{\mu\nu} = - {m^2 \over i\hbar} [\dot x^\mu, \dot x^\nu].
\label{eqn:2.6}
\end{equation}
{}From the Jacobi identity,
\begin{equation}
[\: x^\lambda,[\:\dot x^\mu, \dot x^\nu \:]\:] + [\: \dot x^\mu, [\:
\dot x^\nu,  x^\lambda \:]\:] + [\: \dot x^\nu, [\: x^\lambda , \dot
x^\nu \:]\:] =0
\label{eqn:2.7}
\end{equation}
and (\ref{eqn:2.2}), we show that
\begin{eqnarray}
	[\: x^\lambda, W^{\mu \nu} \:]
	&=&
	- \frac{m^2}{i\hbar} \,
	[\: x^\lambda, [\: \dot x^\mu, \dot x^\nu \:] \:]
	\nonumber
	\\
	&=&
	- \frac{m^2}{i\hbar}
	\Bigl(
	  [\: [\: x^\lambda, \dot x^\mu \:], \dot x^\nu \:]
	+ [\: \dot x^\mu, [\: x^\lambda, \dot x^\nu \:] \:]
	\Bigr)
	\nonumber
	\\
	&=&
	m
	\Bigl(
	  [\: g^{\lambda \mu}, \dot x^\nu \:]
	+ [\: \dot x^\mu, g^{\lambda \nu} \:]
	\Bigr)
	\nonumber
	\\
	&=&
	- i\hbar
	( \partial^\nu g^{\lambda \mu}
	- \partial^\mu g^{\lambda \nu} ).
\label{eqn:2.8}
\end{eqnarray}
So, defining $F^{\mu\nu} = - F^{\nu\mu}$ by
\begin{equation}
F^{\mu\nu} = W^{\mu\nu} - m \langle (\partial^\nu g^{\lambda\mu} -
\partial^\mu g^{\lambda\nu}) \dot x_\lambda \rangle
\label{eqn:2.9}
\end{equation}
where the brackets $ \langle ...  \rangle $ represent Weyl
ordering of the non-commuting operators,
we find from (\ref{eqn:2.8}) and (\ref{eqn:2.9})
\begin{equation}
[x^\sigma, F^{\mu\nu}] = 0,
\label{eqn:2.10}
\end{equation}
which shows that $F^{\mu\nu}$ is independent of $\dot x$.  When
lowering indices, we define
\begin{equation}
\dot x_\mu =  \langle g_{\mu\nu} (x) \dot x^\nu  \rangle \ .
\label{eqn:lower-def}
\end{equation}

With the identity
\begin{equation}
-{m^2 \over i\hbar} [\dot x_{\mu} ,\dot x{_\nu}] = -{m^2 \over
i\hbar} [ \langle g_{\mu\alpha} \dot x^\alpha \rangle ,
 \langle g_{\nu\beta} \dot x^\beta \rangle ] =
 \langle g_{\mu\alpha} g_{\nu\beta} W^{\alpha\beta}  \rangle - m
(\partial_{\mu}  \langle g_{\nu\alpha} \dot x^\alpha  \rangle -
\partial_{\nu} \langle g_{\mu\beta} \dot x^\beta \rangle )
\label{eqn:2.11}
\end{equation}
one sees that
\begin{equation}
F_{\mu\nu} = g_{\mu\alpha} g_{\nu\beta} F^{\alpha\beta}
= -{m^2 \over i\hbar}  [\dot x_\mu ,\dot x_\nu]
\label{eqn:2.12}
\end{equation}
and the Jacobi identity then gives
\begin{equation}
\partial_\mu F_{\nu\rho} + \partial_\nu F_{\rho\mu}+ \partial_\rho
F_{\mu\nu } = 0 .
\label{eqn:2.13}
\end{equation}
Rearranging equation (\ref{eqn:2.5}) and using (\ref{eqn:2.6}),

we see that
\begin{equation}
m[x^\mu ,\ddot x^\nu] = {i\hbar \over m} F^{\mu\nu} + 2i\hbar
 \langle \Gamma^{\nu\lambda\mu} \dot x_\lambda  \rangle
\label{eqn:2.14}
\end{equation}
where
\begin{equation}
\Gamma^{\nu\lambda\mu}= -{1 \over 2} ( \partial^\mu g^{\lambda\nu}
+ \partial^\lambda g^{\mu\nu} - \partial^\nu g^{\lambda\mu})
\label{eqn:2.15}
\end{equation}
is the Levi-Civita connection.  Let us define $ G^\mu$ through the
equation
\begin{equation}
F^\mu = m \ddot x^\mu = G^\mu(x,\dot x,\tau) +
 \langle F^{\mu\nu} \dot x_\nu  \rangle -
m  \langle \Gamma^{\mu\lambda\nu} \dot x_\lambda \dot x_\nu \rangle
 \  .
\label{eqn:2.16}
\end{equation}
Then,
\begin{eqnarray}
[\: x^\lambda, G^\mu \:] &=& [\: x^\lambda, F^\mu \:]
- F^{\mu \nu} [\: x^\lambda, \dot x_\nu \:]
\nonumber \\
&&\mbox{\qquad}+ m \, \Gamma^{\mu \nu \rho} \,
[\: x^\lambda, \dot x_\nu \:] \, \dot x_\rho
+ \Gamma^{\mu \nu \rho} \, \dot x_\nu [\: x^\lambda, \dot x_\rho \:]
\nonumber \\
&=& \frac{i\hbar}{m} \, F^{\lambda \mu}
+ 2 \, i\hbar \,  \langle \Gamma^{\mu \rho \lambda} \dot x_\rho
\rangle
+ \frac{i\hbar}{m} \, F^{\mu \nu} \, \delta^\lambda_{\;\:\nu}
\nonumber \\
&&\mbox{\qquad}- i\hbar \,  \langle \left(
\Gamma^{\mu \nu \rho} \, \delta^\lambda_{\;\:\nu} \, \dot x_\rho
+ \Gamma^{\mu \nu \rho} \, \dot x_\nu \,
\delta^\lambda_{\;\:\rho} \right) \rangle
\nonumber \\
&=& 0,
\label{eqn:2.17}
\end{eqnarray}
so that $G^\mu$ is independent of $\dot x$.  We may then define the
force as
\begin{equation}
G^\mu +  \langle F^{\mu\nu} \dot x_\nu \rangle  =
m [ \ddot x^\mu +
 \langle \Gamma^{\mu\lambda\nu} \dot x_\lambda \dot x_\nu  \rangle
 ]
= m \frac{D \dot x^\mu}{D \tau} \;.
\label{eqn:2.18}
\end{equation}
Since
\begin{equation}
m \ddot x^\mu = m { d \over d \tau}
\langle g^{\mu\nu} \dot x_\nu \rangle ,
\label{eqn:2.19}
\end{equation}
when we lower the index of $G^\mu$ (by (\ref{eqn:2.18}), a tensor)
we find that
\begin{equation}
G_\nu =	g_{\nu \alpha} \, F^\alpha
-  \langle g_{\nu \alpha} \, F^{\alpha \beta} \, \dot x_\beta
\rangle
+ m  \langle g_{\nu \alpha} \, \Gamma^{\alpha \beta \gamma}
\, \dot x_\beta \dot x_\gamma \rangle \ .
\label{eqn:2.20}
\end{equation}
We write the first term on the right hand side of
(\ref{eqn:2.20}) in the form
\begin{eqnarray}
g_{\nu \alpha} \, F^\alpha &=& m \,  \langle g_{\nu \alpha} \,
\ddot x^\alpha \rangle
\nonumber \\
&=& m\, g_{\nu \alpha}\, \frac{d}{d\tau} \langle
g^{\alpha\beta}\dot x_\beta \rangle
\nonumber \\
&=& m \,  \langle g_{\nu \alpha} \, g^{\alpha \beta} \, \ddot x_\beta
+ g_{\nu \alpha} \, \partial^\gamma g^{\alpha \beta}
\, \dot x_\beta \dot x_\gamma
\rangle
\nonumber \\
&=& m \, \ddot x_\nu + m \,  \langle g_{\nu \alpha} \,
\partial^\gamma g^{\alpha \beta} \, \dot x_\beta \dot x_\gamma
\rangle \ .
\label{eqn:2.21}
\end{eqnarray}
Since the indices $\beta$ and $\gamma$ of
$\partial^\gamma g^{\alpha\beta}$ occur in (\ref{eqn:2.21}) in
symmetric combination, we may write
\begin{equation}
\frac12 \, (\partial^\gamma g^{\alpha \beta} +
\partial^\beta  g^{\alpha \gamma} )
= - \Gamma^{\alpha \beta \gamma} +
\frac12 \, \partial^\alpha g^{\beta \gamma}.
\label{eqn:2.22}
\end{equation}
Equations (\ref{eqn:2.20}), (\ref{eqn:2.21}), and (\ref{eqn:2.22})
imply
\begin{eqnarray}
G_\nu &=& m \, \ddot x_\nu +
\frac12 m \,  \langle g_{\nu \alpha} \,
\partial^\alpha g^{\beta \gamma}
\, \dot x_\beta \dot x_\gamma  \rangle
-  \langle g_{\nu \alpha} \, F^{\alpha \beta} \, \dot x_\beta
\rangle
\nonumber \\
&=& m \, \ddot x_\nu +
\frac12 m \,  \langle \partial_\nu g^{\alpha \beta}
\, \dot x_\alpha \dot x_\beta  \rangle
-  \langle F_{\nu \alpha} \, g^{\alpha \beta} \, \dot x_\beta  \rangle
\ .
\label{eqn:2.23}
\end{eqnarray}
Using Equations (\ref{eqn:2.4}), (\ref{eqn:2.12}), and
(\ref{eqn:2.23}) we obtain
\begin{eqnarray}
[\: \dot x_\mu, G_\nu \:] &=&
m \: [\: \dot x_\mu, \ddot x_\nu \:] \:	+ \langle
\frac12 i\hbar \partial_\mu \partial_\nu g^{\alpha \beta}
	\, \dot x_\alpha \dot x_\beta
- \frac{i\hbar}{2m} \partial_\nu g^{\alpha \beta}
( F_{\mu \alpha} \dot x_\beta + \dot x_\alpha F_{\mu \beta} )
\nonumber \\
&&\mbox{\qquad}
- \frac{i\hbar}m \partial_\mu ( F_{\nu \alpha} \,
g^{\alpha \beta} ) \, \dot x_\beta + \frac{i\hbar}{m^2}
\, F_{\nu \alpha} \, g^{\alpha \beta} \, F_{\mu \beta} \rangle
\nonumber\\
&=& m \: [\: \dot x_\mu, \ddot x_\nu \:] \: +
 \langle \frac12 i\hbar \partial_\mu \partial_\nu g^{\alpha \beta}
\, \dot x_\alpha \dot x_\beta - \frac{i\hbar}m
(\partial_\nu g^{\alpha \beta}) F_{\mu \alpha} \, \dot x_\beta
\nonumber\\
&&\mbox{\qquad}
- \frac{i\hbar}m (\partial_\mu g^{\alpha \beta})
F_{\nu \alpha} \, \dot x_\beta 	- \frac{i\hbar}m
\partial_\mu F_{\nu \alpha} g^{\alpha \beta} \,
\dot x_\beta
\nonumber \\
&&\mbox{\qquad}+\frac{i\hbar}{m^2}
\, F_{\nu \alpha} \, g^{\alpha \beta} \, F_{\mu \beta}  \rangle
\ .
\label{eqn:2.24}
\end{eqnarray}
Finally antisymmetrization with respect to the indices
$\mu$ and $\nu$ gives
\begin{eqnarray}
[\: \dot x_\mu, G_\nu \:] - [\: \dot x_\nu, G_\mu \:] &=&
m \: [\: \dot x_\mu, \ddot x_\nu \:] -
[\: \dot x_\nu, \ddot x_\mu \:] - \frac{i\hbar}m \:
 \langle ( \partial_\mu F_{\nu \alpha} - \partial_\nu F_{\mu \alpha} )
g^{\alpha \beta} \, \dot x_\beta  \rangle
\nonumber \\
&=& m \, \frac{d}{d \tau} [\: \dot x_\mu, \dot x_\nu \:]
- \frac{i\hbar}m \:  \langle ( \partial_\mu F_{\nu \alpha}
+ \partial_\nu F_{\alpha \mu} ) \dot x^\alpha \rangle
\nonumber \\
&=& - \frac{i\hbar}m \frac{d}{d \tau} F_{\mu \nu} - \frac{i\hbar}m
 \langle ( \partial_\mu F_{\nu \rho} + \partial_\nu F_{\rho \mu} )
 \dot x^\rho \rangle
\nonumber \\
&=& - \frac{i\hbar}m \: \left[  \langle ( \partial_\rho F_{\mu \nu}
+ \partial_\mu  F_{\nu \rho}
+ \partial_\nu  F_{\rho \mu} ) \dot x^\rho  \rangle +
\partial_\tau F_{\mu \nu} \right] \ .
\label{eqn:2.25}
\end{eqnarray}
Therefore, using (\ref{eqn:2.13}),
\begin{equation}
\partial_{\mu} G_\nu - \partial_{\nu} G_\mu +{ \partial F_{\mu\nu}
\over \partial\tau} = 0.
\label{eqn:2.26}
\end{equation}
Regarding equations (\ref{eqn:2.18}), (\ref{eqn:2.13}), and
(\ref{eqn:2.26}) in the Ehrenfest sense, we may summarize the
classical theory as
\begin{eqnarray}
m \frac{D \dot x^\mu}{D \tau} =
m [ \ddot x^\mu +
\Gamma^{\mu\lambda\nu} \dot x_\lambda \dot x_\nu ] =
G^\mu +  F^{\mu\nu} \dot x_\nu
\label{eqn:sum1}
\\
\partial_\mu F_{\nu\rho} + \partial_\nu F_{\rho\mu}+ \partial_\rho
F_{\mu\nu } = 0
\label{eqn:sum2}
\\
\partial_{\mu} G_\nu - \partial_{\nu} G_\mu +{ \partial F_{\mu\nu}
\over \partial\tau} = 0
\label{eqn:sum3}
\end{eqnarray}
We see that the expressions for the Lorentz force and the
conditions on the fields reduce to equations (\ref{eqn:1.11}),
(\ref{eqn:1.14}), and (\ref{eqn:1.13}) when the metric and the
fields are taken to be $\tau$-independent.

Let us introduce the definitions
\begin{equation}
x^d=\tau \qquad \partial_\tau = \partial_d \qquad
F_{\mu d} = - F_{d \mu} = G_\mu .
\label{eqn:2.27}
\end{equation}
We may then combine (\ref{eqn:2.13}) and (\ref{eqn:2.26}) as
\begin{equation}
\partial_\alpha F_{\beta \gamma} +
\partial_\beta F_{\gamma \alpha} +
\partial_\gamma F_{\alpha \beta} =0
\label{eqn:2.28}
\end{equation}
(for $\alpha,\beta,\gamma$ = $0,\cdots,d$), which shows that the
two form $F$ is closed on the $(d+1)$-dimensional manifold
$(\tau,x)$.  Hence, if for example, this manifold is
contractable, then $F$ is an exact form which can be obtained
as the derivative of some potential with the form $F = dA$.  The
Lorentz force equation becomes
\begin{eqnarray}
m \frac{D \dot x^\mu}{D \tau} &=&
= F^{\mu\nu} (\tau,x) \dot x_\nu \; + G^\mu (\tau,x)
\nonumber \\
&=& F^\mu_\nu (\tau,x) \dot x^\nu \; +
F^{\mu}\ _{d}(\tau,x) \dot x^d
\nonumber \\
&=& F^{\mu}\ _{\beta} (\tau,x) \dot x^\beta .
\label{eqn:2.29}
\end{eqnarray}

Following Dyson and Feynman, we may observe that given equation
(\ref{eqn:2.28}), the two-form $F^{\alpha\beta}$ is determined
if we know functions $\rho$ and $j^\mu$ such that
\begin{eqnarray}
{\cal D}_{\alpha} \; F^{\mu\alpha} &=& j^\mu
\label{eqn:2.34}
\\
{\cal D}_{\alpha} \; F^{d\alpha} &=& \rho.
\label{eqn:2.35}
\end{eqnarray}
where ${\cal D}_{\alpha}$ is the covariant derivative.

Unlike the Newtonian case, the Lorentz covariance of these
expressions is manifest, and we expect that $j^\mu$
transforms as a $d$-vector and $\rho$ transforms as a scalar.
By denoting $\rho=j^d$, equations (\ref{eqn:2.34}) and
(\ref{eqn:2.35}) can be written compactly as
\begin{equation}
{\cal D}_\alpha \; F^{\beta\alpha} = j^\beta
\label{eqn:2.36}
\end{equation}
where, due to the antisymmetry of $F^{\beta\alpha}$, $j^\beta$
is conserved
\begin{equation}
{\cal D}_{\alpha} j^\alpha =0.
\label{eqn:2.37}
\end{equation}
Notice that (\ref{eqn:2.37}) admits a formal $d+1$-dimensional
symmetry (as does the homogeneous field equation (\ref{eqn:2.28})),
owing to
the sum on $\alpha = 0, \cdots , d$.  However, the physical Lorentz
covariance of the matter currents breaks this formal symmetry.

In Section 4, we will return to the equations derived here
formally, and examine their meaning as a covariant canonical
mechanics.

\section{Aspects of the Inverse Problem in the \newline Calculus of
Variations}
\setcounter{equation}{0}

In \cite{H-S}, Hojman and Shepley set out to prove that Feynman's
program of finding the Maxwell theory from equations
(\ref{eqn:1.3}) --- (\ref{eqn:1.5}) without a Lagrangian, was in
principle impossible, because these equations are sufficient to
establish the existence of a unique Lagrangian of electromagnetic
form.  To accomplish this goal, they demonstrate a new connection
between the commutation relations and well-established results
from the inverse problem in the calculus of variations, a theory
which concerns the conditions under which a system of
differential equations may be derived from a variational
principle.  We briefly review elements of this theory and use
these results to derive Tanimura's equations from self-adjointness.

For the situation relevant to Lagrangian mechanics, we consider a
set of ordinary second order differential equations of the form
\begin{equation}
F_k (\tau,q,\dot q,\ddot q) = 0 \qquad \dot q^j = \frac{dq^j}{d\tau}
\qquad \ddot q^j = \frac{d^2q^j}{d\tau^2} \qquad j,k=1,\cdots,n
\ .
\label{eqn:3.1}
\end{equation}
Under variations of the path
\begin{eqnarray}
q(\tau) &\longrightarrow& q(\tau) + dq(\tau)
\nonumber \\
\dot q(\tau) &\longrightarrow& \dot q(\tau) + d \dot q(\tau) =
\dot q(\tau) + \frac{d}{d\tau} dq(\tau)
\nonumber \\
\ddot q(\tau) &\longrightarrow& \ddot q(\tau) + d \ddot q(\tau) =
\ddot q(\tau) + \frac{d^2}{d\tau^2} dq(\tau) \ ,
\label{eqn:basic1}
\end{eqnarray}
the function $F_k (\tau,q,\dot q,\ddot q)$ admits the variational
one-form defined by
\begin{equation}
dF_k = \frac{\partial F_k}{\partial q^j} dq^j +
\frac{\partial F_k}{\partial \dot q^j} d \dot q^j +
\frac{\partial F_k}{\partial \ddot q^j} d \ddot q^j
\ ,
\label{eqn:basic2}
\end{equation}
and the variational two-form,
\begin{equation}
dq^k dF_k = \frac{\partial F_k}{\partial q^j} dq^k \wedge dq^j +
\frac{\partial F_k}{\partial \dot q^j} dq^k \wedge d \dot q^j +
\frac{\partial F_k}{\partial \ddot q^j} dq^k \wedge d \ddot q^j
\label{eqn:basic3}
\end{equation}
where the $3n$ path variations $(dq^k,d\dot q^k,d\ddot q^k)$
for $k=1,\cdots,n$ are understood to be linearly independent.
The system of differential equations
$F_k(\tau,q,\dot q,\ddot q)$ is called self-adjoint if
there exists a two-form
$\Omega_2(dq,d \dot q)$ such that for all admissible
variations of the path,
\begin{equation}
dq^k dF_k(dq) = \frac{d}{d\tau} \Omega_2(dq,d \dot q) \ .
\label{eqn:basic4}
\end{equation}
Through integration by parts, one may show \cite{Santilli}
that such a two-form exists and is unique up to an additive
constant, if and only if
\begin{eqnarray}
\frac{\partial F_i}{\partial \ddot q^k} &=&
\frac{\partial F_k}{\partial \ddot q^i}
\label{eqn:3.2} \\
\frac{\partial F_i}{\partial \dot q^k} +
\frac{\partial F_k}{\partial \dot q^i} &=& \frac{d}{d\tau}
\left[ \frac{\partial F_i}{\partial \ddot q^k} +
\frac{\partial F_k}{\partial \ddot q^i} \right]
\label{eqn:3.3} \\
\frac{\partial F_i}{\partial q^k} - \frac{\partial F_k}{\partial q^i}
&=& \frac{1}{2} \frac{d}{d\tau}
\left[ \frac{\partial F_i}{\partial \dot q^k} -
\frac{\partial F_k}{\partial \dot q^i} \right] \ .
\label{eqn:3.4}
\end{eqnarray}
Equations (\ref{eqn:3.2}) -- (\ref{eqn:3.4}) are known as the
Helmholtz conditions \cite{Helmholtz,Darboux}.  Introducing the
notation
\begin{equation}
\delta = dq^k_\beta \frac{\partial}{\partial q^k_\beta} \qquad
q^k_\beta = \left(\frac{d}{d\tau}\right)^\beta q^k
\qquad \beta=0,1,2 \ ,
\label{eqn:basic5}
\end{equation}
it follows that
\begin{equation}
\delta^2 = dq^k_\beta \wedge dq^l_\alpha
\frac{\partial^2}{\partial q^k_\beta \partial q^l_\alpha} =0 \ ,
\label{eqn:basic6}
\end{equation}
which permits the equivalence of a set of self-adjointness
differential equations to a Lagrangian formulation to be easily
demonstrated \cite{Dedecker}.  Given the Lagrangian $L$,
\begin{equation}
\delta L = \frac{\partial L}{\partial q^k} dq^k +
\frac{\partial L}{\partial \dot q^k} d\dot q^k =
\left[ -\frac{d}{d\tau} \frac{\partial L}{\partial \dot q^k} +
\frac{\partial L}{\partial q^k} \right] dq^k +
\frac{d}{d\tau} \left(\frac{\partial L}{\partial \dot q^k}
dq^k \right) = F_k dq^k + \frac{d}{d\tau} \Omega_1
\label{eqn:basic7}
\end{equation}
Therefore,
\begin{equation}
\delta^2 = 0 \Longrightarrow - dq^k \delta F_k +
\frac{d}{d\tau} \delta \Omega_1 =
- dq^k \delta F_k + \frac{d}{d\tau}\Omega_2 = 0
\label{eqn:basic8}
\end{equation}
which demonstrates self-adjointness.
Conversely, if $F_k$ is self-adjoint, then
$dq^k \delta F_k - \frac{d}{d\tau}\Omega_2 = 0$ and since
$\delta^2=0$,
\begin{equation}
\frac{d}{d\tau}\Omega_2 = \delta \frac{d}{d\tau}\Omega_1 \ .
\label{eqn:basic9}
\end{equation}
Therefore,
\begin{equation}
0=dq^k \delta F_k - \frac{d}{d\tau}\Omega_2 =
\delta (dq^k F_k - \frac{d}{d\tau}\Omega_1) = \delta L
\label{eqn:basic}
\end{equation}
and one obtains the differential equations $F_k = 0$ by variation
of $L$ under $\tau$-integration.

For the second order equations considered here, it also follows
\cite{Santilli} from self-adjointness that the most general form
of $F_k$ is
\begin{equation}
F_k (\tau,q,\dot q,\ddot q) = A_{kj} (\tau,q,\dot q) \ddot q^j
+ B_k (\tau,q,\dot q).
\label{eqn:3.5}
\end{equation}
To see this, notice that $F_k$ is independent of $d^3 q^i /dt^3$,
so that the right hand side of (\ref{eqn:3.3})
must be independent of $\ddot q^i$.
Inserting (\ref{eqn:3.5}) into (\ref{eqn:3.2}) -- (\ref{eqn:3.4}),
one finds the Helmholtz conditions on $A_{kj}$ and $B_k$
\begin{eqnarray}
A_{ij} = A_{ji}\qquad&&\qquad \frac{\partial A_{ij}}{\partial \dot q^k}
= \frac{\partial A_{kj}}{\partial \dot q^i}
\label{eqn:3.6} \\
\frac{\partial B_i}{\partial \dot q^j} +
\frac{\partial B_j}{\partial \dot q^i} &=&
2 \left[ \frac{\partial}{\partial \tau} +
\dot q^k \frac{\partial}{\partial q^k} \right] A_{ij}
\label{eqn:3.7} \\
\frac{\partial B_i}{\partial q^j} -
\frac{\partial B_j}{\partial q^i} &=&
\frac12 \left[ \frac{\partial}{\partial \tau} +
\dot q^k \frac{\partial}{\partial q^k} \right]
\left( \frac{\partial B_i}{\partial \dot q^j} -
\frac{\partial B_j}{\partial \dot q^i} \right)
\label{eqn:3.8}
\end{eqnarray}
along with the useful identity
\begin{equation}
\frac{\partial A_{ij}}{\partial q^k} -
\frac{\partial A_{kj}}{\partial q^i} =
\frac12 \frac{\partial }{\partial \dot q^j}
\left( \frac{\partial B_i}{\partial \dot q^k} -
\frac{\partial B_k}{\partial \dot q^i} \right).
\label{eqn:3.9}
\end{equation}

In the domain of invertibilty of the $A_{jk}$, one can write
(\ref{eqn:3.5}) as
\begin{equation}
F_k (\tau,q,\dot q,\ddot q) = A_{kj} (\tau,q,\dot q) [ \ddot q^j
- f^j  ] ,
\label{eqn:3.10}
\end{equation}
where
\begin{equation}
f^j (\tau,q,\dot q) = -(A^{-1})^{jk} B_k .
\label{eqn:3.11}
\end{equation}
Inserting $B_k = -A_{kj} f^j$ into (\ref{eqn:3.7}) and
(\ref{eqn:3.8}), the Helmholtz conditions for the form
(\ref{eqn:3.10}) become
\begin{eqnarray}
A_{ij} = A_{ji} \qquad && \qquad
\frac{\partial A_{ij}}{\partial \dot q^k} =
\frac{\partial A_{kj}}{\partial \dot q^i}
\label{eqn:3.12} \\
\frac{D}{D\tau} A_{ij} &=& - \frac{1}{2} \left[
A_{ik} \frac{\partial f^k}{\partial \dot q^j} +
A_{jk} \frac{\partial f^k}{\partial \dot q^i} \right]
\label{eqn:3.13} \\
\frac{1}{2} \frac{D}{D\tau} \left[
A_{ik} \frac{\partial f^k}{\partial \dot q^j} -
A_{jk} \frac{\partial f^k}{\partial \dot q^i} \right]
&=& A_{ik} \frac{\partial f^k}{\partial q^j} -
A_{jk} \frac{\partial f^k}{\partial q^i}
\label{eqn:3.14}
\end{eqnarray}
where
\begin{equation}
\frac{D}{D\tau} = \frac{\partial }{\partial \tau} +
\dot q^k \frac{\partial }{\partial q^k} +
f^k \frac{\partial }{\partial \dot q^k}
\label{eqn:3.15}
\end{equation}
is the total time derivative subject to the constraint
\begin{equation}
\ddot q^k - f^k (\tau,q,\dot q) =0.
\label{eqn:3.16}
\end{equation}
The identity (\ref{eqn:3.9}) becomes
\begin{equation}
\frac{\partial A_{ij}}{\partial q^k} -
\frac{\partial A_{kj}}{\partial q^i}
=
-\frac{1}{2} \frac{\partial}{\partial \dot q^j} \left[
\frac{\partial }{\partial \dot q^k}(A_{in}f^n) -
\frac{\partial }{\partial \dot q^i}(A_{kn}f^n) \right] .
\label{eqn:3.17}
\end{equation}

Within the domain of applicability of the inverse function
theorem, (\ref{eqn:3.16}) is equivalent to (\ref{eqn:3.5}), and
the Helmholtz conditions become the necessary and sufficient
conditions for the existence of an integrating factor $A_{jk}$
such that
\begin{equation}
F_k = A_{kj} (\tau,q,\dot q) [ \ddot q^j - f^j  ] = \frac{d}{d\tau}
\left( \frac{\partial L}{\partial \dot q^k} \right)
-\frac{\partial L}{\partial q^k} .
\label{eqn:3.18}
\end{equation}
The Helmholtz conditions for this form have been rederived in an
elegant way in terms of the Lie derivative along $f^k$
\cite{Henneaux}.

In \cite{H-S}, Hojman and Shepley prove that given the quantum
mechanical commutation relations
\begin{equation}
[ X^i (\tau) , \dot X^j (\tau) ] = i \hbar G^{ij} \ ,
\label{eqn:3.19}
\end{equation}
the classical function
\begin{equation}
g^{ij} = \lim_{\hbar \rightarrow 0} G^{ij}
\label{eqn:3.20}
\end{equation}
has an inverse
\begin{equation}
\omega_{ij} = (g^{-1})_{ij}
\label{eqn:3.21}
\end{equation}
which satisfies the Helmholtz conditions
(\ref{eqn:3.12}) -- (\ref{eqn:3.14}).
Therefore, the assumption of commutation relations, (\ref{eqn:1.3})
and (\ref{eqn:1.4}) [or (\ref{eqn:1.8}) and (\ref{eqn:1.9})], is
sufficiently strong to establish the existence of an equivalent
Lagrangian for the classical problem associated with the quantum
commutators.  For the Newtonian case, in which
$g^{ij} = \delta^{ij}$ and $\tau \rightarrow t$,
it is shown in \cite{H-S,Hughes}
that the Helmholtz conditions lead to the Lagrangian
(\ref{eqn:1.6}), with field equations (\ref{eqn:1.7}).  In
\cite{Santilli}, Santilli discusses the classical case, applying
(\ref{eqn:3.2}) -- (\ref{eqn:3.4}) to (\ref{eqn:3.10}) for the
case $A_{ij} = \delta_{ij}$, and arrives at (\ref{eqn:1.1}) and
(\ref{eqn:1.2}) without explicitly writing the Lagrangian.

We now adapt Santilli's argument to the type of curved
space discussed in Section 2.  Starting with the Helmholtz
conditions and the metric $g_{\mu\nu}(x)$,
we will be led to the equations derived by Tanimura.
We take the function $A_{\mu\nu} = g_{\mu\nu} (x)$
in equation
(\ref{eqn:3.20}) to be a Riemannian metric independent of $\dot x$,
so that equations
(\ref{eqn:3.12}) are satisfied automatically.  Since $g_{\mu\nu}$
does not depend on $\dot x^\mu$, equation (\ref{eqn:3.13}) becomes
\begin{equation}
\frac{D}{D \tau} g_{\mu\nu} =
\dot x^\sigma \frac{\partial}{\partial x^\sigma} g_{\mu\nu} =
-\frac{1}{2} \left[
\frac{\partial f_\mu}{\partial \dot x^\nu} -
\frac{\partial f_\nu}{\partial \dot x^\mu} \right]
\label{eqn:3.22}
\end{equation}
and equation (\ref{eqn:3.17}) becomes
\begin{equation}
-\frac{1}{2} \frac{\partial}{\partial \dot x^\nu} \left[
\frac{\partial f_\mu}{\partial \dot x^\sigma} -
\frac{\partial f_\sigma}{\partial \dot x^\mu} \right]
= \frac{\partial g_{\mu\nu}}{\partial x^\sigma} -
\frac{\partial g_{\sigma\nu}}{\partial x^\mu} .
\label{eqn:3.23}
\end{equation}
Acting on (\ref{eqn:3.22}) with $\partial / \partial \dot x^\sigma$
and exchanging ($\nu \leftrightarrow \sigma$), we obtain
\begin{equation}
g_{\mu\sigma,\nu} = -\frac{1}{2} \left[
\frac{\partial^2 f_\mu}{\partial \dot x^\sigma \partial \dot x^\nu} +
\frac{\partial^2 f_\sigma}{\partial \dot x^\mu \partial \dot x^\nu}
\right]
\label{eqn:3.24}
\end{equation}
where $g_{\mu\sigma,\nu} = \partial g_{\mu\sigma} / \partial x^\nu$.
Combining (\ref{eqn:3.23}) and (\ref{eqn:3.24}), we find
\begin{equation}
\frac{1}{2}
\frac{\partial^2 f_\mu}{\partial \dot x^\sigma \partial \dot x^\nu} =
-\frac{1}{2} (g_{\mu\nu,\sigma} + g_{\mu\sigma,\nu} -
g_{\sigma\nu,\mu}) = -\Gamma_{\mu\sigma\nu}
\label{eqn:3.25}
\end{equation}
where $\Gamma_{\mu\sigma\nu}$ is defined as in (\ref{eqn:2.15}).
{}From (\ref{eqn:3.25}) we see that the most general expression for
$f_\mu (\tau,x,\dot x)$ is
\begin{equation}
f_\mu = - \Gamma_{\mu\nu\sigma} \dot x^\nu \dot x^\sigma -
\rho_{\mu\nu}(\tau,x)\dot x^\nu - \sigma_\mu (\tau,x).
\label{eqn:3.26}
\end{equation}
Now from (\ref{eqn:3.22}) we find
\begin{equation}
\dot x^\sigma \frac{\partial g_{\mu\nu}}{\partial x^\sigma} =
\frac{1}{2} \left[ 2 \Gamma_{\mu\nu\sigma} \dot x^\sigma +
2 \Gamma_{\nu\mu\sigma} \dot x^\sigma + \rho_{\mu\nu} +
\rho_{\nu\mu} \right] .
\label{eqn:3.27}
\end{equation}
Using
\begin{equation}
(\Gamma_{\mu\nu\sigma} + \Gamma_{\nu\mu\sigma} )
\dot x^\sigma = g_{\mu\nu,\sigma} \dot x^\sigma
\label{eqn:3.28}
\end{equation}
we find that all terms except for those in $\rho_{\mu\nu}$ cancel,
so that
\begin{equation}
0= \rho_{\mu\nu} + \rho_{\nu\mu} .
\label{eqn:3.29}
\end{equation}
We now apply equation (\ref{eqn:3.14}) which becomes
\begin{eqnarray}
\frac{1}{2} \frac{D}{D\tau} \left[
g_{\mu\sigma} \frac{\partial f^\sigma}{\partial \dot x^\nu} -
g_{\nu\sigma} \frac{\partial f^\sigma}{\partial \dot x^\mu} \right]
&=& g_{\mu\sigma} \frac{\partial f^\sigma}{\partial x^\nu} -
g_{\nu\sigma} \frac{\partial f^\sigma}{\partial x^\mu}
\nonumber \\
\frac{1}{2} \frac{D}{D\tau} \left[
\frac{\partial f_\mu}{\partial \dot x^\nu} -
\frac{\partial f_\nu}{\partial \dot x^\mu} \right]
&=& f_{\mu , \nu} - f_{\nu , \mu} - g_{\mu\sigma , \nu} f^\sigma
+ g_{\nu\sigma , \mu} f^\sigma .
\label{eqn:3.30}
\end{eqnarray}
Using (\ref{eqn:3.26}) to expand the left hand side,
\begin{eqnarray}
\frac{1}{2} \frac{D}{D\tau} \left[
\frac{\partial f_\mu}{\partial \dot x^\nu} -
\frac{\partial f_\nu}{\partial \dot x^\mu} \right]
&=& - \frac{1}{2} \frac{D}{D\tau} \biggl[
\frac{\partial }{\partial \dot x^\nu} \Bigl(
\Gamma_{\mu\lambda\sigma} \dot x^\lambda \dot x^\sigma +
\rho_{\mu\lambda}(\tau,x)\dot x^\lambda
\nonumber \\
&&\mbox{\qquad\qquad\qquad}
+ \sigma_\mu (\tau,x)
\Bigr) - (\mu \leftrightarrow \nu) \biggr]
\nonumber \\
&=& - \frac{1}{2} \frac{D}{D\tau} \biggl[
2(\Gamma_{\mu\nu\lambda} - \Gamma_{\nu\mu\lambda} ) \dot x^\lambda
+ \rho_{\mu\nu} -
\rho_{\nu\mu} \biggr]
\nonumber \\
&=& - \left(\frac{\partial }{\partial \tau} +
\dot x^\sigma \frac{\partial }{\partial x^\sigma} +
f^\sigma \frac{\partial }{\partial \dot x^\sigma} \right)
\left[ (g_{\mu\lambda,\nu} - g_{\nu\lambda,\mu}) \dot x^\lambda +
\rho_{\mu\nu} \right]
\nonumber \\
&=& - (g_{\mu\sigma,\nu} - g_{\nu\sigma,\mu}) f^\sigma
- \rho_{\mu\nu,\tau}
\nonumber \\&&\mbox{\qquad\qquad}
- \dot x^\lambda \dot x^\sigma
(g_{\mu\lambda,\nu\sigma} - g_{\nu\lambda,\mu\sigma}) +
\dot x^\lambda \rho_{\mu\nu,\lambda}
\label{eqn:3.31}
\end{eqnarray}
where $\rho_{\mu\nu,\tau} = \partial \rho_{\mu\nu} /
\partial \tau$, and we have used
\begin{eqnarray}
2(\Gamma_{\mu\nu\lambda} - \Gamma_{\nu\mu\lambda} ) \dot x^\lambda
&=& \dot x^\lambda (-g_{\nu\lambda,\mu} + g_{\mu\lambda,\nu} +
g_{\nu\mu,\lambda} + g_{\mu\lambda,\nu} - g_{\nu\lambda,\mu} -
g_{\mu\nu,\lambda})
\nonumber \\
&=& 2 \dot x^\lambda ( g_{\mu\lambda,\nu} - g_{\nu\lambda,\mu}).
\label{eqn:3.32}
\end{eqnarray}
{}From (\ref{eqn:3.26}), we have
\begin{eqnarray}
f_{\mu,\nu} &=&
- \biggl[ \Gamma_{\mu\lambda\sigma} \dot x^\lambda \dot x^\sigma +
\rho_{\mu\lambda}(\tau,x) \dot x^\lambda +
\sigma_\mu (\tau,x) \biggr]_{,\nu}
\nonumber \\
&=& - \biggl[ \Gamma_{\mu\lambda\sigma,\nu}
\dot x^\lambda \dot x^\sigma
+ \rho_{\mu\lambda,\nu}\dot x^\lambda +
\sigma_{\mu,\nu}\biggr]
\label{eqn:3.33}
\end{eqnarray}
so that the right hand side of (\ref{eqn:3.30}) is
\begin{eqnarray}
f_{\mu , \nu} - f_{\nu , \mu} && \mbox{\hspace{-.4 in}}
- g_{\mu\sigma , \nu} f^\sigma + g_{\nu\sigma , \mu} f^\sigma
= \mbox{\hspace*{1.5 in}}
\nonumber \\
&=& - \biggl[ (\Gamma_{\mu\lambda\sigma,\nu} -
\Gamma_{\nu\lambda\sigma,\mu} )
\dot x^\lambda \dot x^\sigma
+ (\rho_{\mu\lambda,\nu}-  \rho_{\nu\lambda,\mu})\dot x^\lambda
+ \sigma_{\mu,\nu} - \sigma_{\nu,\mu} \biggr]
\nonumber \\&&\mbox{\qquad}
- (g_{\mu\sigma , \nu} - g_{\nu\sigma , \mu}) f^\sigma.
\label{eqn:3.34}
\end{eqnarray}
Now, equating (\ref{eqn:3.31}) and (\ref{eqn:3.34}) and canceling
common terms, we are left with
\begin{equation}
\frac{\partial\rho_{\mu\nu}}{\partial \tau} + \dot x^\lambda
\rho_{\mu\nu,\lambda} = \dot x^\lambda (\rho_{\mu\lambda,\nu} -
\rho_{\nu\lambda,\mu}) + \sigma_{\mu,\nu} - \sigma_{\nu,\mu}.
\label{eqn:3.35}
\end{equation}
Since the $\dot x^\lambda$ are arbitrary, we have the two
expressions
\begin{eqnarray}
\frac{\partial\rho_{\mu\nu}}{\partial \tau}
= \frac{\partial \sigma_\mu}{\partial x^\nu} -
\frac{\partial \sigma_\nu}{\partial x^\mu}
\label{eqn:3.36} \\
\partial_\lambda \rho_{\mu\nu} + \partial_\mu \rho_{\nu\lambda} +
\partial_\nu \rho_{\lambda\mu} =0
\label{eqn:3.37}
\end{eqnarray}
Comparing (\ref{eqn:3.37}) with (\ref{eqn:2.13}), (\ref{eqn:3.36})
with (\ref{eqn:2.26}), and (\ref{eqn:3.26}) with
(\ref{eqn:2.16}), we see that we may identify
\begin{equation}
F_{\mu\nu} = -\rho_{\mu\nu} \qquad{\rm and}
\qquad G_\mu = -\sigma_\mu \ .
\label{eqn:3.38}
\end{equation}
This identification demonstrates explicitly that Tanimura's
equations (generalized to permit explicit $\tau$ dependence of
the fields and metric) are simply the conditions on the most
general velocity dependent forces which may be obtained from a
variational principle.

To make the connection with the off-shell electromagnetic theory,
we now derive the Lagrangian in flat Minkowski space.  For this
case, where
\begin{equation}
A_{\mu\nu} = g_{\mu\nu} \longrightarrow \eta_{\mu\nu}
= {\rm diag} (1,-1,\cdots,-1)
\label{eqn:3.39}
\end{equation}
Tanimura's equations reduce to
\begin{eqnarray}
\partial_\mu F_{\nu\rho} + \partial_\nu F_{\rho\mu}+ \partial_\rho
F_{\mu\nu } = 0
\nonumber \\
\partial_{\mu} G_\nu - \partial_{\nu} G_\mu +{ \partial F_{\mu\nu}
\over \partial\tau} = 0
\nonumber \\
m \ddot x^\mu =  G^\mu(\tau,x) + F^{\mu\nu}(\tau,x) \dot x_\nu
\label{eqn:3.40}
\end{eqnarray}
Following \cite{Santilli}, we observe that (\ref{eqn:3.18})
implies
\begin{eqnarray}
\eta_{\mu\nu} [m \ \ddot x^\nu - f^\nu  ] &=& \frac{d}{d\tau}
\left( \frac{\partial L}{\partial \dot x^\mu} \right)
-\frac{\partial L}{\partial x^\mu}
\nonumber \\
&=&
\frac{\partial^2 L}{\partial \dot x^\mu \partial \dot x^\nu}
\ddot x^\nu
+ \frac{\partial^2 L}{\partial \dot x^\mu \partial x^\nu}\dot x^\nu
+ \frac{\partial^2 L}{\partial \dot x^\mu \partial \tau}
- \frac{\partial L}{\partial x^\mu}
\label{eqn:3.41}
\end{eqnarray}
so that
\begin{eqnarray}
m \ \eta_{\mu\nu} =
\frac{\partial^2 L}{\partial \dot x^\mu \partial \dot x^\nu}
\label{eqn:3.42} \\
- \eta_{\mu\nu} f^\nu =
\frac{\partial^2 L}{\partial \dot x^\mu \partial x^\nu}\dot x^\nu
+ \frac{\partial^2 L}{\partial \dot x^\mu \partial \tau}
- \frac{\partial L}{\partial x^\mu}
\label{eqn:3.43}
\end{eqnarray}
The solution to the equation $\partial^2 L_{\rm kinetic} /
\partial \dot x^\mu \partial \dot x^\nu = m\eta_{\mu\nu}$ is
unique up to terms of the type which may be absorbed in
(\ref{eqn:3.43}).  Therefore, we see that $L$ may consist of the
quadratic term integrated from (\ref{eqn:3.42}),
plus terms at most linear in $\dot x^\mu$.  So we may write
\begin{equation}
L = \frac12 m \ \dot x^\mu \dot x_\mu + A_\mu(\tau,x) \dot x^\mu +
\phi (\tau,x) \ .
\label{eqn:3.44}
\end{equation}
Applying (\ref{eqn:3.43}) to $L$ yields the Lorentz force
if we identify the fields $A_\mu$ and $\phi$ as the potentials for the
field strengths
\begin{eqnarray}
F_{\mu\nu} &=& \partial_\mu A_\nu - \partial_\nu A_\mu
\label{eqn:f.42} \\
G_\mu &=& \partial_\mu \phi - \partial_\tau A_\mu \ .
\label{eqn:f.43}
\end{eqnarray}
We may recover the inhomogeneous field equations (\ref{eqn:2.34})
and (\ref{eqn:2.35})
by introducing to the action the kinetic terms
\begin{equation}
\int d^4 x d\tau \left[
\frac14 F_{\mu\nu}F^{\mu\nu} \pm \frac12 G_\mu G^\mu \right] \ .
\label{eqn:f.44}
\end{equation}
Applying the Euler-Lagrange equations to the action which includes these
terms leads to the source equations, with the identification of the
classical currents as
\begin{eqnarray}
j^\mu (\tau,y)&=& \dot x^\mu (\tau) \delta^4 \Bigl( y-x(\tau) \Bigr)
\label{eqn:f.45} \\
\rho (\tau,y)&=& \delta^4 \Bigl( y-x(\tau) \Bigr) \ .
\label{eqn:f.46}
\end{eqnarray}
If the fields are considered $\tau$-independent {\it a priori},
as by Tanimura, the currents are
\begin{eqnarray}
j^\mu (y)&=& \int d\tau \, \dot x^\mu (\tau) \,
\delta^4 \Bigl( y-x(\tau) \Bigr)
\label{eqn:f.47} \\
\rho (y)&=& \int d\tau \, \delta^4 \Bigl( y-x(\tau) \Bigr) \ .
\label{eqn:f.48}
\end{eqnarray}
Equations (\ref{eqn:2.35}) and (\ref{eqn:f.48}) indicate that the
$\tau$-independent $G^\mu$ field in Tanimura's formulation must have
a non-trivial form related to the path of the particle event, and
also, influence the particle motions through the Lorentz force, with no
coupling to $F^{\mu\nu}$ which could control the interaction.
In the next section,
we arrive at (\ref{eqn:3.44}) by a different argument and
examine the dynamical system which it describes.

\section{Implications for Gauge Theory
in \newline Covariant Quantum Mechanics}

\setcounter{equation}{0}

In this section, we show the consequences of the analysis
discussed above for a Lorentz and gauge covariant quantum
mechanics.  We first review the structure of such a theory.

In 1951,
Schwinger \cite{Schwinger} represented the Green's functions of
the Dirac field as a parametric integral and formally
transformed the Dirac problem into a dynamical one, in which
the integration parameter acts as a proper time according to
which a ``Hamiltonian'' operator generates the evolution of the
system through spacetime.  Applying Schwinger's method to the
Klein-Gordon equation, one obtains an equation for the Green's
function (we take $\hbar =1 $ in the following)
\begin{equation}
G=\frac{1}{(p-eA)^{2} +m^{2}}
\label{eqn:4.2}
\end{equation}
given by
\begin{equation}
G(x,x')=\langle x|G|x'\rangle =i \int_{0}^{\infty} ds e^{-i m^{2} s}
\langle x|e^{-i (p-eA)^{2} s}|x'\rangle  .
\label{eqn:4.3}
\end{equation}
The function
\begin{equation}
G(x,x';s)=\langle x(s)|x'(0)\rangle = \langle x|e^{-i (p-eA)^{2} s}|x'\rangle
\label{eqn:4.4}
\end{equation}
satisfies
\begin{equation}
i \frac{\partial}{\partial s}\langle x(s)|x'(0)\rangle =
(p-eA)^{2}\langle x(s)|x'(0)\rangle =K\langle x(s)|x'(0)\rangle
\label{eqn:4.5}
\end{equation}
with the boundary condition
\begin{equation}
\lim_{s \rightarrow 0} \langle x(s)|x'(0)\rangle  = \delta ^{4} (x-x').
\label{eqn:4.6}
\end{equation}
DeWitt \cite{DeWitt} regarded
(\ref{eqn:4.5}) as defining the Green's function for the
Schr\"odinger equation
\begin{equation}
i \frac{\partial}{\partial s}\psi_{s}(x)=K\psi_{s}(x)=
(p-eA)^{2}\psi_{s}(x) \ ,
\label{eqn:4.7}
\end{equation}
an equation which Stueckelberg \cite{Stueckelberg} had also written
as the basis for a covariant quantum mechanical formalism which
includes the description of pair annihilation.
Feynman \cite{Feynman} wrote equation (\ref{eqn:4.7}) in order
to obtain the path
integral for the Klein-Gordon equation and regarded the
integration of the Green's function with the weight
$e^{-im^{2}s}$, as the requirement that asymptotic solutions be on
mass-shell.  The usual Feynman propagator $\Delta_{F} (x-x')$
emerges naturally from the assumption that $G(x,x';s)=0$
whenever $s<0$.  Thus $\Delta_{F}$ corresponds to a $G(x,x';s)$
which is causal in the classical sense of no response before
stimulus (before, in the sense of $s$).  Schwinger employed the
proper time method as a way to exploit the techniques of
nonrelativistic mechanics in relativistic quantum theory, and
found that it provided a natural approach to perturbation theory
and regularization.

In 1973, Horwitz and Piron \cite{H-P} constructed a canonical
formalism for the relativistic classical and quantum mechanics of
many particles.  In order to formulate a generalized Hamilton's
principle, they introduce a Lorentz invariant evolution parameter
$\tau$, which they call the historical time and regard as
corresponding to the ordering relation of successive events in
spacetime.  For a one-particle system, the equations of motion are
\begin{equation}
\frac{dx^{\mu}}{d\tau} =\frac{\partial K}{\partial p_{\mu}}
\qquad\qquad
\frac{dp^{\mu}}{d\tau} =-\frac{\partial K}{\partial x_{\mu}}
\label{eqn:4.8}
\end{equation}
where $\mu,\nu = 0,1,2,3$ and $K(x^{\mu},p^{\mu})$ a Lorentz
scalar.  Taking $K=p^{2}/2M$ leads to the usual description of
the relativistic motion of a free particle:
\begin{equation}
\dot x^0 = \frac{p^0}{M} \qquad
\dot x^i = \frac{p^i}{M} \qquad
p^\mu = \rm constant
\label{eqn:4.9}
\end{equation}
and so
\begin{equation}
\frac{dx^i}{dt} = \frac{p^i}{p^0} \qquad \qquad
\dot x^2 = -\frac{m^2}{M^2} = {\rm constant}
\label{eqn:4.10}
\end{equation}
with $m/M$ scaling $\tau$ to the proper time.  This
formalism leads naturally to a Schr\"odinger equation for the
quantum theory, which is identical to the Stueckelberg
equation (\ref{eqn:4.7}) for the Klein-Gordon problem (for
\mbox{$2M=1$}).

Saad, Horwitz, and Arshansky later argued \cite{Saad} that the
local gauge covariance of the Schr\"odinger equation should
include transformations which depend on $\tau$, as well as on
the spacetime coordinates.  This requirement of full gauge
covariance leads to a theory of five gauge compensation fields,
since gauge transformations are functions on the five dimensional
space $(\tau,x)$.
Under local gauge transformations of the form
\begin{equation}
\psi(x,\tau) \rightarrow e^{i e_{0} \Lambda (x,\tau)} \psi (x,\tau)
\label{eqn:4.11}
\end{equation}
the equation
\begin{equation}
-(i \partial_{\tau} -e_{0}a_{4})\psi(x,\tau)
=\frac{1}{2M}(p^{\mu}-e_{0}a^{\mu})(p_{\mu}-
e_{0}a_{\mu})\psi(x,\tau)
\label{eqn:4.12}
\end{equation}
is covariant, when the compensation fields transform as
\begin{equation}
a_{\mu}(x,\tau) \rightarrow
a_{\mu}(x,\tau)+\partial_{\mu}\Lambda(x,\tau) \qquad
a_{4}(x,\tau) \rightarrow a_{4}(x,\tau) + \partial_{\tau}
\Lambda(x,\tau) .
\label{eqn:4.14}
\end{equation}
This Schr\"odinger equation (\ref{eqn:4.12}) leads to the five
dimensional conserved current [compare with equation
(\ref{eqn:2.37})]
\begin{equation}
\partial_{\mu}j^{\mu}+\partial_{\tau}j^{4}=0
\label{eqn:4.15}
\end{equation}
where
\begin{equation}
j^{4} = |\psi(x,\tau)|^{2} \qquad
j^{\mu} = \frac{-i }{2M}(\psi^{*}(\partial^{\mu}-i
e_{0}a^{\mu})\psi - \psi(\partial^{\mu}-i e_{0}a^{\mu})\psi^{*}).
\label{eqn:4.16}
\end{equation}
In analogy to nonrelativistic quantum mechanics the squared
amplitude of the wave function may be interpreted as the
probability of finding an event at $(\tau,x)$.  Equation
(\ref{eqn:4.15}) may be written as
$\partial_{\alpha}j^{\alpha}=0$, with $\alpha=0,1,2,3,4$.

{}From (\ref{eqn:4.12}) we can write the classical Hamiltonian as
\begin{equation}
K=\frac{1}{2M}(p^{\mu}-e_{0}a^{\mu})(p_{\mu}-
e_{0}a_{\mu})-e_{0}a_{4}
\label{eqn:h-cl}
\end{equation}
and using (\ref{eqn:4.8}) we find
\begin{equation}
M \ \dot x^\mu = (p^{\mu}-e_{0}a^{\mu})
\label{eqn:p-cl}
\end{equation}
which enables us to write the classical Lagrangian,
\begin{eqnarray}
L &=& \dot x^\mu p_\mu - K
\nonumber \\
&=& \frac12 M \dot x^\mu \dot x_\mu + e_0 \dot x^\mu a_\mu
+ e_{0}a_{4}.
\label{eqn:l-cl}
\end{eqnarray}
Comparing (\ref{eqn:l-cl}) with (\ref{eqn:3.44}), we see that
identifying the
scalar potential $\phi (\tau,x)$ with $e_{0}a_{4}$ and the vector
potential $A_\mu(\tau,x)$ with $e_0 a_\mu$, this Lagrangian
describes
the same theory which we derived from the Helmholtz conditions and
from Tanimura's commutation relations
(\ref{eqn:1.8}) and (\ref{eqn:1.9}).  We may find the Lorentz force
\cite{emlf} by applying the Euler-Lagrange equations to
(\ref{eqn:l-cl}), which in the notation of (\ref{eqn:2.29}),
i.e., $\alpha,\beta = 0,1,2,3,4$,  is
\begin{equation}
M \: \ddot x^\mu = f^\mu_{\:\;\;\nu} \dot x^\nu + f^\mu_{\:\;\;4}
= f^\mu_{\:\;\;\alpha}(x,\tau) \, \dot x^\alpha .
\label{eqn:4.18}
\end{equation}
where
\begin{equation}
f^{\mu\nu} = \partial^\mu a^\nu - \partial^\nu a^\mu
\qquad f^\mu_{\:\;\;4} = \partial^\mu a_4 - \partial_\tau a^\mu
\label{eqn:f-def}
\end{equation}
Since $f^{\mu\nu}$ and $f^\mu_{\:\;\;4}$ are defined in terms of
potentials, the field equations, (\ref{eqn:2.13}) and
(\ref{eqn:2.26}) [or (\ref{eqn:3.37}) and (\ref{eqn:3.36})]
follow by definition.

Comparing (\ref{eqn:4.18}) with (\ref{eqn:1.11}) and
(\ref{eqn:3.26}), we identify $f_{\mu 4}$ with the
``new'' field
$G_\mu$ found by Tanimura (and the field $\sigma_\mu$ which
emerges from the derivation based on the Helmholtz conditions).
This justifies our claim in Section 1 that the appearance of the
``new'' scalar degree of gauge freedom in Tanimura's derivation
corresponds to invariance of the theory under gauge
transformations which depend upon the evolution parameter $\tau$.

We observe that the four equations (\ref{eqn:4.18}) imply
\cite{emlf} that
\begin{equation}
\frac{d}{d\tau} (\frac{1}{2} M \dot x^2) = M \dot x^\mu
\ddot x_\mu = \dot x^\mu (f_{\mu 4} + f_{\mu\nu} \dot
x^\nu ) = \dot x^\mu f_{\mu 4}
\label{eqn:4.n1}
\end{equation}
By replacing $f_{\mu 4}$ with $-f_{4\mu}$ and using $f_{44}\equiv 0$,
we can write equation (\ref{eqn:4.n1}) as
\begin{equation}
\frac{d}{d\tau} (-\frac{1}{2} M \dot x^2) = f_{4\alpha}
\dot x^\alpha
\label{eqn:4.n2}
\end{equation}
which formally becomes the ``fifth'' component of the Lorentz
force law.  If we define the $4 \oplus 1$ vector
\begin{equation}
U^\alpha = ( -\frac{1}{2} \dot x^2 , \dot x^\mu )
\label{eqn:4.n3}
\end{equation}
then (\ref{eqn:4.n1}) and (\ref{eqn:4.n2}) can be written as
\begin{equation}
M \: \frac{d}{d\tau} U^\alpha = f^\alpha_{\:\;\;\beta} \:
\dot x^\beta .
\label{eqn:4.n4}
\end{equation}

We are now in a position to address the mass-shell condition
$\dot x^2={\rm constant}$ discussed in Section 1.
Since the right hand side of (\ref{eqn:4.18}) is the most general
expression which may appear as a Lorentz force, we may
conclude from (\ref{eqn:4.n2}) that the conditions for the
dynamical (as opposed to
asymptotic) conservation of $\dot x^2 = {\rm constant}$, are
\begin{equation}
G_{\mu} = f_{4\mu} = 0  \qquad {\rm and}  \qquad
\partial_\tau f^{\mu\nu}=0
\label{eqn:conditions}
\end{equation}
The second condition follows from (\ref{eqn:2.26}) for $G^{\mu}=0$;
that is $f^{\mu\nu}$ must be a
$\tau$-static field.  Thus, we see that the most general
interaction which preserves the proper time constraint is of
conventional Maxwell type, as employed by Schwinger \cite{Schwinger}
in his original use of the proper time method.  In this way, we
may make precise the claim that the proper time constraint
suppresses the ``new'' gauge degree of freedom.
Notice that the fields in these expressions are explicitly
$\tau$-dependent.

The action, given by
\begin{equation}
S= \int d\tau \ \frac12 M \dot x^\mu \dot x_\mu +
e_0 \, \dot x^\mu \, a_\mu (\tau,x) + e_{0} \, a_{4} (\tau,x)
\label{eqn:4.act}
\end{equation}
is clearly not reparameterization invariant, corresponding to
the absence of the mass-shell constraint.  But, under the
conditions (\ref{eqn:conditions}), the Lagrangian has no
explicit $\tau$-dependence (see (\ref{eqn:f-def})),
and so the Hamiltonian is conserved.
{}From (\ref{eqn:h-cl}) and (\ref{eqn:p-cl}), we see that under those
circumstances, $K \rightarrow M \dot x^2 /2$.  For those
interactions
which respect the proper time relation and keep particles on
mass-shell dynamically, the $\tau$-derivative of the proper time
is essentially the constant of motion conserved by Noether's
theorem for the $\tau$-translation symmetry.  Thus, the
mass-shell relation has the status, classically, of a conservation
law rather than a constraint.

When we add as the dynamical term for the gauge field,
$(\lambda/4) f_{\alpha \beta}f^{\alpha \beta}$
where $\lambda$ is a dimensional constant,
the equations for the field are found to be
\begin{equation}
\partial_{\beta} f^{\alpha \beta}
=\frac{e_{0}}{\lambda}j^{\alpha}=ej^{\alpha}
\label{eqn:4.19}
\end{equation}
\begin{equation}
\epsilon^{\alpha \beta \gamma \delta
\epsilon}\partial_{\alpha}f_{\beta \gamma}=0
\label{eqn:ho}
\end{equation}
where $f_{\alpha\beta}=
\partial_{\alpha}a_{\beta}-\partial_{\beta}a_{\alpha}$, and
\begin{eqnarray}
j^\mu (\tau,y)&=& \dot x^\mu (\tau) \delta^4 \Bigl( y-x(\tau) \Bigr)
\label{eqn:c.45} \\
j^4 (\tau,y) &=& \rho (\tau,y)= \delta^4 \Bigl( y-x(\tau) \Bigr) \ .
\label{eqn:c.46}
\end{eqnarray}
We identify $e_0 / \lambda $ as the dimensionless Maxwell
charge (it follows from (\ref{eqn:4.22}) below that $e_0$ has
dimension of length).  The currents (\ref{eqn:c.45}) and
(\ref{eqn:c.46}) have the form of (\ref{eqn:f.45}) and
(\ref{eqn:f.46}), so that (\ref{eqn:4.19}) is the flat space
equivalent of (\ref{eqn:2.36}).  Similarly,
(\ref{eqn:ho}) is the homogeneous
equation (\ref{eqn:2.28}) and (\ref{eqn:3.36}--\ref{eqn:3.37}).

The Lagrangian for the free electromagnetic field has a
formal five dimensional symmetry.  Analysis
of the resulting wave equation \cite{Saad}
for the sourceless case
shows that the symmetry group of the equations can be either
O(3,2) or O(4,1), depending on the signature of the $\tau$ index.
Since $U^\alpha$ (see (\ref{eqn:4.n3})) transforms as a
Lorentz scalar plus Lorentz 4-vector, rather than as a vector of
O(3,2) or O(4,1), the presence of sources breaks this formal
symmetry to O(3,1).  This
situation is analogous to the nonrelativistic case discussed in
Section 1, in which the
homogeneous field equations (\ref{eqn:1.2}) may be regarded as
O(3,1) covariant, while the source dynamics (\ref{eqn:1.4}), may
be seen as having a Galilean symmetry; in a consistent theory of
sources and fields, only the common O(3) symmetry survives.

Since the 4-vector part of the current in (\ref{eqn:4.16}) is not
conserved by itself, it may not be the source for the Maxwell field.
However, integration of (\ref{eqn:4.16}) over $\tau$,
with appropriate boundary conditions,
leads to $\partial_{\mu}J^{\mu}=0$, where
\begin{equation}
J^{\mu}(x)=\int_{-\infty}^{\infty} d\tau j^{\mu}(x,\tau)
\label{eqn:cat-j}
\end{equation}
so that we may identify $J^{\mu}$ as the source of the Maxwell
field.  Under appropriate boundary conditions, integration of
(\ref{eqn:4.19}) over $\tau$ implies
\begin{equation}
\partial_{\nu}F^{\mu \nu}=eJ^{\mu}
\label{eqn:4.21}
\end{equation}
\[\epsilon^{\mu \nu \rho \lambda }\partial_{\mu}F_{\nu \rho}=0\]
where
\begin{equation}
F^{\mu \nu}(x)=\int_{-\infty}^{\infty} d\tau f^{\mu \nu}(x,\tau)
\label{eqn:4.22}
\end{equation}
\[A^{\mu}(x)=\int_{-\infty}^{\infty} d\tau a^{\mu}(x,\tau)\]
so that $a^{\alpha}(x,\tau)$ has been called the pre-Maxwell field.

In the pre-Maxwell theory, interactions take place between
events in spacetime rather than between worldlines.
Each event, occurring at $\tau$, induces a current density in
spacetime which disperses for large $\tau$, and the continuity
equation (\ref{eqn:4.15}) states that these current densities
evolve as the event density $j^{4}$ progresses through
spacetime as a function of $\tau$.  As noted above, if
$j^{4}\rightarrow 0$ as $|\tau| \rightarrow \infty$ (pointwise
in spacetime), then the integral of
$j^{\mu}$ over $\tau$ may be identified with the Maxwell current.
This integration has been called
concatenation \cite{concat} and provides the link between the
event along a worldline and the notion of a particle, whose
support is the entire worldline.  Concatenation is evidently
related to the integration performed in the Stueckelberg
theory, and following Feynman's interpretation, places the
electromagnetic field on the zero mass-shell.  The Maxwell
theory has the character of an {\it equilibrium limit} of the
microscopic pre-Maxwell theory.  For further discussion and
applications see
\cite{bound,shnerb,frastai,qft,ruther,selrul,zeeman} and
references contained therein.

\section{Non-Abelian Gauge Theory}
\setcounter{equation}{0}

It was shown in Section 4 that while the mass-shell condition may
not be maintained as a constraint on the phase space, the
quantity $\dot x^\mu \dot x_\mu$ is a constant of the motion when
the ``new'' gauge field $G_{\mu} = F_{4\mu}$ vanishes and
$F^{\mu\nu}$ is $\tau$-independent.  Under these conditions, the
Lagrangian has no explicit $\tau$-dependence and the conserved
Hamiltonian is precisely $\dot x^\mu \dot x_\mu$.  We remark
briefly on the case of a non-Abelian gauge field, in which a
$\tau$-dependent quantity appears in the Lagrangian, without
changing these conditions on the fields.

In \cite{Lee}, C. R. Lee employed Feynman's method to derive
equations of motion for a particle interacting with a classical
non-Abelian gauge field, in the form originally given by Wong
\cite{Wong}.  By studying the Heisenberg equations of motion for
the Hamiltonian of the Dirac equation in the presence of an SU(2)
gauge field, Wong formulated the following structure:
\begin{eqnarray}
m \ddot \xi_\mu &=& g {\bf f}_{\mu\nu} \cdot {\bf I}(\tau)
\ \dot \xi^\nu
\label{eqn:w1}\\
\dot {\bf I} &=& - g {\bf b}_\mu \times {\bf I} \ \dot \xi^\mu
\label{eqn:w2}\\
{\bf f}_{\mu\nu} &=& \partial_\mu {\bf b}_\nu -
\partial_\nu {\bf b}_\mu + g {\bf b}_\mu \times {\bf b}_\nu
\label{eqn:w3}\\
\partial^\mu {\bf f}_{\mu\nu} + g {\bf b}^\mu \times {\bf f}_{\mu\nu}
&=& - {\bf j}_\nu
\label{eqn:w4}\\
{\bf b}^\mu = b_{a\mu} I^a \qquad {\bf f}_{\mu\nu} &=& f_{a\mu\nu} I^a
\qquad [I^a,I^b] = i\hbar \varepsilon^{abc} I_c.
\label{eqn:w5}
\end{eqnarray}
where $\xi^\mu (\tau)$ is the particle world line operator as
parameterized by the Lorentz invariant scalar $\tau$ and where
the $I^a(\tau)$ are an operator representation of the generators
of a non-Abelian gauge group.  By virtue of (\ref{eqn:w3}), one
has the inhomogeneous equation
\begin{equation}
{\cal D}_\mu {\bf f}_{\nu \rho} + {\cal D}_\nu  {\bf f}_{\rho \mu} +
{\cal D}_\rho {\bf f}_{\mu \nu} = 0,
\label{eqn:w6}
\end{equation}
where the covariant derivative is
\begin{equation}
( {\cal D}_\mu {\bf f}_{\mu\nu} )_a = \partial_\mu f_{a \mu\nu}
- \varepsilon_a^{\; bc} b_{b \mu} f_{c\mu\nu}.
\label{eqn:w7}
\end{equation}

Lee \cite{Lee} followed Feynman's method, supplementing assumptions
(\ref{eqn:1.3}) --- (\ref{eqn:1.5}) with the relations
(\ref{eqn:w5}) and
\begin{equation}
[x_i , I^a(t)] = 0 \qquad
\dot {\bf I} + g {\bf b}_i \times {\bf I} \ \dot x^i = 0
\label{eqn:lee1}
\end{equation}
for $i=1,2,3$, and in analogy to Feynman's derivation,
arrived at the Newtonian equivalent of Wong's equations.

Tanimura \cite{Tanimura} generalized Lee's derivation to
$d$-dimensional flat Minkowski space and a general gauge group by
supplementing equations (\ref{eqn:2.1}) and (\ref{eqn:2.2}) with
\begin{equation}
[\: I^a, I^b \:] = i\hbar  \, f_c^{\; ab} I^c
\label{eqn:5.1}
\end{equation}
\begin{equation}
[\: x^\mu, I^a \:] = 0
\label{eqn:5.2}
\end{equation}
\begin{equation}
m \: \ddot x^\mu = F^\mu (x,\dot x,I) = F_a^\mu (x,\dot x) \, I^a
\label{eqn:5.3}
\end{equation}
and
\begin{equation}
\dot I^a =  f_c^{\;ab} A_{b \mu}(x) \, \dot x^\mu I^c
\label{eqn:5.4}
\end{equation}
The results of his derivation are
\begin{equation}
m \: \ddot x^\mu = G_a^\mu(x) \, I^a +
F^\mu_{a \; \nu}(x) \, I^a \dot x^\nu
\label{eqn:5.5}
\end{equation}
where the fields satisfy
\begin{equation}
( {\cal D}_\mu G_\nu - {\cal D}_\nu G_\mu )_a = 0
\label{eqn:5.6}
\end{equation}
\begin{equation}
({\cal D}_\mu  F_{\nu \rho} + 	{\cal D}_\nu  F_{\rho \mu} +
{\cal D}_\rho F_{\mu \nu} )_a = 0,
\label{eqn:5.7}
\end{equation}
and where the form of the covariant derivative is
\begin{equation}
( {\cal D}_\mu F_{\nu \rho} )_a = \partial_\mu F_{a \nu \rho}
- f_a^{\; bc} A_{b \mu} F_{c \, \nu \rho} .
\label{eqn:5.8}
\end{equation}
The field strength $F_{\nu \rho}$ is related to $A_{b \mu}$ through
\begin{equation}
f_c^{\; ab} \Bigl( F_{a \mu \nu} -
( \partial_\mu A_{a \nu} - \partial_\nu A_{a \mu}
- f_a^{\; de} A_{d \mu} A_{e \nu} ) \Bigr) = 0 \ .
\label{eqn:5.9}
\end{equation}

The non-Abelian theory may be examined from the point of view of
Section 3, by generalizing the Helmholtz conditions to take
account of classical non-Abelian gauge fields according to
Wong's formulation.  To achieve this, we associate with
variations $dq$ of the path $q(\tau)$, a variation $dI^a$ of the
generators $I^a$, which may be understood as the variation of the
orientation of the tangent space under $q(\tau)
\rightarrow q(\tau) + dq(\tau)$.  The explicit form of this
variation follows from (\ref{eqn:5.4}): for small $d\tau$,
\begin{equation}
dI^a =  f_c^{\;ab} [A_{b \mu}(\tau,x) \, dx^\mu +
\phi_b(\tau,x) d\tau ] I^c
\label{eqn:5.10}
\end{equation}
where we have allowed an explicit $\tau$-dependence for the gauge
field, and have included a Lorentz scalar gauge field $\phi_a$,
in analogy
with the Abelian case.  Now, the quantity ${\bf M}=M_a I^a$
undergoes the variation of the path
\begin{equation}
(\tau,x) \longrightarrow (\tau + d\tau ,x + dx )
\label{eqn:variation}
\end{equation}
according to
\begin{eqnarray}
d {\bf M} &=& (d M_a) I^a + M_a (d I^a)
\nonumber \\
&=& \left(\frac{\partial M_a}{\partial \tau} d\tau +
\frac{\partial M_a}{\partial x^\mu} dx^\mu +
\frac{\partial M_a}{\partial \dot x^\mu} d \dot x^\mu +
\frac{\partial M_a}{\partial \ddot x^\mu} d \ddot x^\mu\right)I^a +
M_a [f_c^{\;ab} A_{b \mu} \, dx^\mu +
\phi_b d\tau ]I^c
\nonumber \\
&=& \left[\frac{\partial M_a}{\partial \tau}-
f_a^{\;bc} \phi_b M_c \right]I^a d\tau +
\left[\frac{\partial M_a}{\partial x^\mu} -
f_a^{\;bc} A_{b \mu}M_c  \right]I^a  dx^\mu +
\frac{\partial M_a}{\partial \dot x^\mu}I^a d \dot x^\mu +
\frac{\partial M_a}{\partial \ddot x^\mu}I^a d \ddot x^\mu
\nonumber \\
&=& {\cal D}_\tau {\bf M}  d\tau +
{\cal D}_\mu {\bf M}  dx^\mu +
\frac{\partial {\bf M} }{\partial \dot x^\mu} d \dot x^\mu +
\frac{\partial {\bf M} }{\partial \ddot x^\mu} d \ddot x^\mu
\label{eqn:5.11}
\end{eqnarray}
in which the spacetime part of the covariant derivative ${\cal D}_\mu$
has the form of (\ref{eqn:5.8}), and a similar covariant derivative
for the $\tau$ component appears which contains $\phi_a$.
Now, the entire structure presented in Section 3 follows with the
replacements
\begin{equation}
\frac{\partial}{\partial x^\mu} \longrightarrow {\cal D}_\mu
\qquad
\frac{\partial}{\partial \tau} \longrightarrow {\cal D}_\tau \ ,
\label{eqn:5.12}
\end{equation}
so that the Helmholtz conditions become
\begin{eqnarray}
A_{\mu\nu} = A_{\nu\mu} \qquad && \qquad
\frac{\partial A_{\mu\nu}}{\partial \dot x^\sigma} =
\frac{\partial A_{\sigma\nu}}{\partial \dot x^\mu}
\label{eqn:h.12} \\
\frac{D}{D\tau} A_{\mu\nu} &=& - \frac{1}{2} \left[
A_{\mu\sigma} \frac{\partial f^\sigma}{\partial \dot x^\nu} +
A_{\nu\sigma} \frac{\partial f^\sigma}{\partial \dot x^\mu} \right]
\label{eqn:h.13} \\
\frac{1}{2} \frac{D}{D\tau} \left[
A_{\mu\sigma} \frac{\partial f^\sigma}{\partial \dot x^\nu} -
A_{\nu\sigma} \frac{\partial f^\sigma}{\partial \dot x^\mu} \right]
&=& A_{\mu\sigma} {\cal D}_\nu f^\sigma -
A_{\nu\sigma} {\cal D}_\mu f^\sigma
\label{eqn:h.14}
\end{eqnarray}
where
\begin{equation}
\frac{D}{D\tau} = {\cal D}_\tau +
\dot x^\sigma {\cal D}_\sigma +
f^\sigma \frac{\partial }{\partial \dot x^\sigma}
\label{eqn:h.15}
\end{equation}
is the total $\tau$ derivative subject to
\begin{equation}
\ddot x_\mu - f_{a\mu} (\tau,x,\dot x) I^a =0 \ .
\label{eqn:h.16}
\end{equation}
Since Hojman and Shepley's argument \cite{H-S} relates only to the
commutation relations among the coordinates, not to the structure
of the forces, their result carries over unchanged.

We now apply equations (\ref{eqn:h.12}) --- (\ref{eqn:h.16}) to
the case of flat spacetime.  Since $A_{\mu\nu} = g_{\mu\nu} =
\eta_{\mu\nu}$
is constant, (\ref{eqn:h.12}) is trivially satisfied and
(\ref{eqn:h.13}) becomes
\begin{equation}
\frac{\partial f_\mu}{\partial \dot x^\nu} +
\frac{\partial f_\nu}{\partial \dot x^\mu} =0
\qquad \Longrightarrow \qquad
\frac{\partial^2 f_\mu}
{\partial \dot x^\nu \partial \dot x^\lambda} +
\frac{\partial^2 f_\nu}
{\partial \dot x^\mu\partial \dot x^\lambda} =0
\ .
\label{eqn:5.13}
\end{equation}
Recalling the identity (\ref{eqn:3.17}), we may also write (since
the metric carries no group indices)
\begin{equation}
\frac{\partial^2 f_\mu}
{\partial \dot x^\nu \partial \dot x^\lambda} -
\frac{\partial^2 f_\nu}
{\partial \dot x^\mu\partial \dot x^\lambda} =0
\ ,
\label{eqn:5.14}
\end{equation}
so that
\begin{equation}
\frac{\partial^2 f_\mu}
{\partial \dot x^\nu \partial \dot x^\lambda}=0
\ .
\label{eqn:5.15}
\end{equation}
Therefore, the most general form of $f_\mu$ is
\begin{equation}
f_\mu = f_{\mu\nu} (\tau,x) \dot x^\nu + g_\mu (\tau,x)
\ ,
\label{eqn:5.16}
\end{equation}
where (\ref{eqn:5.13}) requires that $f_{\mu\nu} + f_{\nu\mu} =0$.
Finally, applying (\ref{eqn:h.14}) we find
\begin{eqnarray}
\frac{1}{2} \frac{D}{D\tau} \left[
\frac{\partial f_\mu }{\partial \dot x^\nu} -
\frac{\partial f_\nu}{\partial \dot x^\mu} \right]
&=& {\cal D}_\nu f_\mu - {\cal D}_\mu f_\nu
\nonumber \\
\frac{1}{2} \frac{D}{D\tau} [f_{\mu\nu} - f_{\nu\mu} ] &=&
{\cal D}_\nu f_{\mu\lambda}\dot x^\lambda +
{\cal D}_\nu g_\mu -
{\cal D}_\mu f_{\nu\lambda}\dot x^\lambda +
{\cal D}_\mu g_\nu
\nonumber \\
({\cal D}_\tau + \dot x^\lambda {\cal D}_\lambda) f_{\mu\nu} &=&
\dot x^\lambda ({\cal D}_\nu f_{\mu\lambda} -
{\cal D}_\mu f_{\nu\lambda}) +
{\cal D}_\nu g_\mu -{\cal D}_\mu g_\nu \ .
\label{eqn:5.17}
\end{eqnarray}
Since $\dot x^\mu$ is arbitrary, we find that
\begin{eqnarray}
{\cal D}_\lambda f_{\mu\nu} +
{\cal D}_\mu f_{\nu\lambda} +
{\cal D}_\nu f_{\lambda\mu} &=& 0
\label{eqn:5.18a} \\
{\cal D}_\tau f_{\mu\nu} + {\cal D}_\mu g_\nu -
{\cal D}_\nu g_\mu &=& 0 \ .
\label{eqn:5.18b}
\end{eqnarray}
Now, in analogy to the Abelian case, we may write
\begin{equation}
L = \frac12 m \ \dot x^\mu \dot x_\mu + A_{a\mu}(\tau,x)I^a(\tau)
\dot x^\mu + \phi_a(\tau,x)I^a(\tau) \ .
\label{eqn:5.19}
\end{equation}
Applying the Euler-Lagrange equations to (\ref{eqn:5.19}), we
obtain
\begin{eqnarray}
\frac{d}{d\tau} \left[ m \dot x_\mu + A_{a\mu}I^a\right]
&=& \frac{\partial}{\partial x^\mu} [A_{a\nu}I^a
\dot x^\nu + \phi_aI^a]
\nonumber \\
m \ddot x_\mu +
\frac{\partial A_{a\mu}}{\partial \tau} I^a +
\frac{\partial A_{a\mu}}{\partial x^\nu} \dot x^\nu I^a
+ A_{a\mu} \dot I^a
&=& \frac{\partial A_{a\nu}}{\partial x^\mu}\dot x^\nu
I^a + \frac{\partial \phi_a}{\partial x^\mu}I^a
\label{eqn:5.20}
\end{eqnarray}
Rearranging terms and using (\ref{eqn:5.10}) to express $\dot
I^a$, we find
\begin{eqnarray}
m \ddot x_\mu &=&\left[ \left(
\frac{\partial A_{a\nu}}{\partial x^\mu}\dot x^\nu -
\frac{\partial A_{a\mu}}{\partial x^\nu} \dot x^\nu
\right) I^a - A_{a\mu} \dot I^a\right] +
\frac{\partial \phi_a}{\partial x^\mu}I^a -
\frac{\partial A_{a\mu}}{\partial \tau} I^a
\nonumber \\
&=&\left[ \left(
\frac{\partial A_{a\nu}}{\partial x^\mu}\dot x^\nu -
\frac{\partial A_{a\mu}}{\partial x^\nu} \dot x^\nu
\right) I^a - A_{a\mu} f_c^{\;ab} (A_{b \nu} \dot x^\nu + \phi_b) I^c
\right] +
\frac{\partial \phi_a}{\partial x^\mu}I^a -
\frac{\partial A_{a\mu}}{\partial \tau} I^a
\nonumber \\
&=&\left[
\frac{\partial A_{a\nu}}{\partial x^\mu}-
\frac{\partial A_{a\mu}}{\partial x^\nu} +
f_a^{\;bc}  A_{a\mu} A_{b \nu} \right] \dot x^\nu I^a
+\left[\frac{\partial \phi_a}{\partial x^\mu} -
\frac{\partial A_{a\mu}}{\partial \tau}+
f_a^{\;bc} A_{a\mu} \phi_b \right] I^a \ .
\label{eqn:5.21}
\end{eqnarray}
Comparing (\ref{eqn:5.21}) with (\ref{eqn:5.16}), we may express
the field strengths in terms of the potentials as
\begin{eqnarray}
f_{\mu\nu} &=& \left[
\frac{\partial A_{a\nu}}{\partial x^\mu}-
\frac{\partial A_{a\mu}}{\partial x^\nu} +
f_a^{\;bc}  A_{a\mu} A_{b \nu} \right] \dot x^\nu I^a
\nonumber \\
g_\mu &=& \left[\frac{\partial \phi_a}{\partial x^\mu} -
\frac{\partial A_{a\mu}}{\partial \tau}+
f_a^{\;bc} A_{a\mu} \phi_b \right] I^a  \ ,
\label{eqn:5.22}
\end{eqnarray}
from which it follows that (\ref{eqn:5.18a}) and (\ref{eqn:5.18b})
are satisfied.  We remark that since Tanimura did not include a
Lorentz scalar potential and his fields were assumed to be
$\tau$-independent, there is no potential in his formulation from
which the field $G_\mu$ in (\ref{eqn:5.5}) and (\ref{eqn:5.6})
could be derived.  On the other hand, by the argument of Hojman
and Shepley, equations (\ref{eqn:5.5}) are equivalent to the
Lagrangian given in (\ref{eqn:5.19}), so that the scalar
potential must be present to obtain a non-zero $G_\mu$.  Unlike
the Abelian case, the potentials appear in the covariant
derivative, and so the ``new'' gauge potential will mix with all
quantities whose covariant derivatives are calculated.

As in (\ref{eqn:2.27}), we may introduce the definitions
\begin{equation}
x^d=\tau \qquad \partial_\tau = \partial_d \qquad
f_{\mu d} = - f_{d \mu} = g_\mu .
\label{eqn:na.27}
\end{equation}
We may then combine (\ref{eqn:5.18a}) and (\ref{eqn:5.18b}) as
\begin{equation}
\partial_\alpha f_{\beta \gamma} +
\partial_\beta f_{\gamma \alpha} +
\partial_\gamma f_{\alpha \beta} =0
\label{eqn:na.28}
\end{equation}
(for $\alpha,\beta,\gamma$ = $0,\cdots,d$).  The
Lorentz force equation becomes
\begin{eqnarray}
m \ddot x^\mu &=&
= f^{\mu\nu}_a \dot x_\nu I^a \; + g^\mu_a I^a
\nonumber \\
&=& f^{\mu\nu}_a I^a \dot x_\nu \; +
f^{\mu}_{a \; d} I^a \dot x^d
\nonumber \\
&=& f^{\mu}_{a \beta} \dot x^\beta .
\label{eqn:na.29}
\end{eqnarray}
where
\begin{equation}
f_{\alpha\beta} = \left[
\frac{\partial A_{a\beta}}{\partial x^\alpha}-
\frac{\partial A_{a\alpha}}{\partial x^\beta} +
f_a^{\;bc}  A_{a\alpha} A_{b \beta} \right] I^a
\label{eqn:na.30}
\end{equation}
recovers the usual relationship of the field strength tensor
to the non-Abelian potential.

We finally examine the conservation of $\dot x^\mu \dot x_\mu$.
In the non-Abelian case, the mass-shell condition becomes
(compare with (\ref{eqn:4.n1}))
\begin{equation}
\frac{d}{d\tau} (\frac{1}{2} m \dot x^2) = m \dot x^\mu
\ddot x_\mu = \dot x^\mu I^a (g_{a \mu} + f_{a \mu\nu} \dot
x^\nu ) = \dot x^\mu g_{a \mu} I^a = 0
\label{eqn:m.13}
\end{equation}
which implies
\begin{equation}
g_{a \mu} = 0  \qquad {\rm and}  \qquad
\partial_\tau f^{a \mu\nu}=0
\label{eqn:m.14}
\end{equation}
where the second expression follows from (\ref{eqn:5.18b}).
Notice that $I^a (\tau)$ introduces a $\tau$-dependence
which is present in the Lagrangian even when the fields are
$\tau$-independent.  But one may easily compute the Hamiltonian
from (\ref{eqn:5.19}) as
\begin{equation}
K = \dot x^\mu  \frac{\partial L}{\partial x^\mu} - L =
\frac{1}{2} m \dot x^\mu \dot x_\mu - \phi_a I^a
\label{eqn:m.15}
\end{equation}
so that (\ref{eqn:m.13}) tells us that the Hamiltonian is
conserved under the conditions (\ref{eqn:m.14}), in apparent
contradiction to the explicit $\tau$-dependence of the Lagrangian.
We may check, however, by explicit calculation that
\begin{equation}
\frac{\partial L}{\partial \tau} = A_{a \mu} \dot x^\mu \dot I^a
+ \phi_a \dot I^a =
f^{abc} A_{a \mu} A_{b \nu} \dot x^\mu \dot x^\nu I^c
+ \phi_a \dot I^a = \phi_a \dot I^a.
\label{eqn:m.16}
\end{equation}
Thus, despite the explicit appearance of $\tau$ in the
Lagrangian, the structure of the non-Abelian field guarantees
that $\partial L / \partial \tau =0$ in the absence of the
``new'' gauge potential $\phi_a$,
leading to the preservation of the mass-shell as conservation of
the Hamiltonian, as in the Abelian case.

\section{Conclusion}
\setcounter{equation}{0}

In Feynman's 1948 derivation and in the powerful technique of
Hojman and Shepley, one sees that the form of the commutation
relations between position and velocity (defined as a derivative
with respect to an independent time parameter) determines the most
general form of the forces which may act on those phase space
variables, and that these forces are of a gauge type.  Since the
commutation relations determine the form of the Lagrangian to be
a scalar with respect to the group which preserves the metric,
and since the evolution parameter $\tau$ is an independent variable,
the Hamiltonian will also be
a scalar and will generate translations of this parameter.
The gauge group may include $\tau$-dependent gauge
transformations, requiring a conjugate compensation field.  The
source-free equations for the gauge fields then admit a
formal spacetime symmetry which is larger than the symmetry group
of original phase space, in which the parameter $\tau$ plays the
role of a coordinate.  In the case of Feynman's
assumption of Newtonian mechanics, the appearance of the fourth
gauge field $A_0$ compensates for $t$-dependent gauge
transformations, and the homogeneous field equations are formally
consistent with a 4-dimensional symmetry, which could be O(3,1).
Similarly, in the case of Tanimura's assumption of $2d$
independent phase space variables with an O($d$-1,1) symmetry, the
appearance of the $d$+1$^{\rm st}$ gauge field compensates for
$\tau$-dependent gauge transformations, and the homogeneous field
equations are formally consistent with a $d$+1-dimensional
symmetry, possibly O($d$,1) or O($d$-1,2).  From the Lorentz force
law for the O($d$-1,1) theory, one may find the $d$+1$^{\rm st}$
expression which explicitly relates the ``new'' field with the
exchange of mass between the field and the sources.  Similarly,
in the O(3) case derived by Feynman, one may derive, from the
three independent components of the Lorentz force law, a fourth
expression which relates the $t$-dependent {\bf E}-field to the
exchange of scalar energy between the field and the sources.

For the case of the O(3,1) theory, one has a means of arriving at
the Maxwell theory of electrodynamics from commutation relations.
Deriving the Maxwell theory in this manner provides a clear
picture of the way in which the usual on-shell dynamics is a
proper restriction of the general off-shell theory.

%
%

\end{document}